# MRI Images, Brain Lesions and Deep Learning

Darwin Castillo[1,2,4*], Vasudevan Lakshminarayanan[2,3], M.J. Rodríguez-Álvarez [4]

[1]*Departamento de Química y Ciencias Exactas, Sección Fisicoquímica y Matemáticas, Universidad Técnica Particular de Loja, Postal Code 11-01-608, Loja, Ecuador*
[2]*Theoretical and Experimental Epistemology Lab, School of Optometry and Vision Science, University of Waterloo, Ontario, Canada*
[3] *Departments of Physics, Electrical and Computer Engineering and Systems Design Engineering, University of Waterloo, Ontario, Canada*
[4]*Instituto de Instrumentación para Imagen Molecular (i3M) Universitat Politècnica de València – Consejo Superior de Investigaciones Científicas (CSIC), Valencia, Spain*

# Abstract

Medical brain image analysis is a necessary step in the Computers Assisted /Aided Diagnosis (CAD) systems. Advancements in both hardware and software in the past few years have led to improved segmentation and classification of various diseases. In the present work, we review the published literature on systems and algorithms that allow for classification, identification, and detection of White Matter Hyperintensities (WMHs) of brain MRI images specifically in cases of ischemic stroke and demyelinating diseases. For the selection criteria, we used the bibliometric networks. Out of a total of 140 documents we selected 38 articles that deal with the main objectives of this study. Based on the analysis and discussion of the revised documents, there is constant growth in the research and proposal of new models of deep learning to achieve the highest accuracy and reliability of the segmentation of ischemic and demyelinating lesions. Models with indicators (Dice Score, DSC: 0.99) were found, however with little practical application due to the uses of small datasets and lack of reproducibility. Therefore, the main conclusion is to establish multidisciplinary research groups to overcome the gap between CAD developments and their complete utilization in the clinical environment.

**Corresponding author:** dpcastillo@utpl.edu.ec, phone +593 07 370 1444; ext.3204, Universidad Técnica Particular de Loja, Postal Code 11-01-608, Loja, Ecuador

# Graphical Abstract

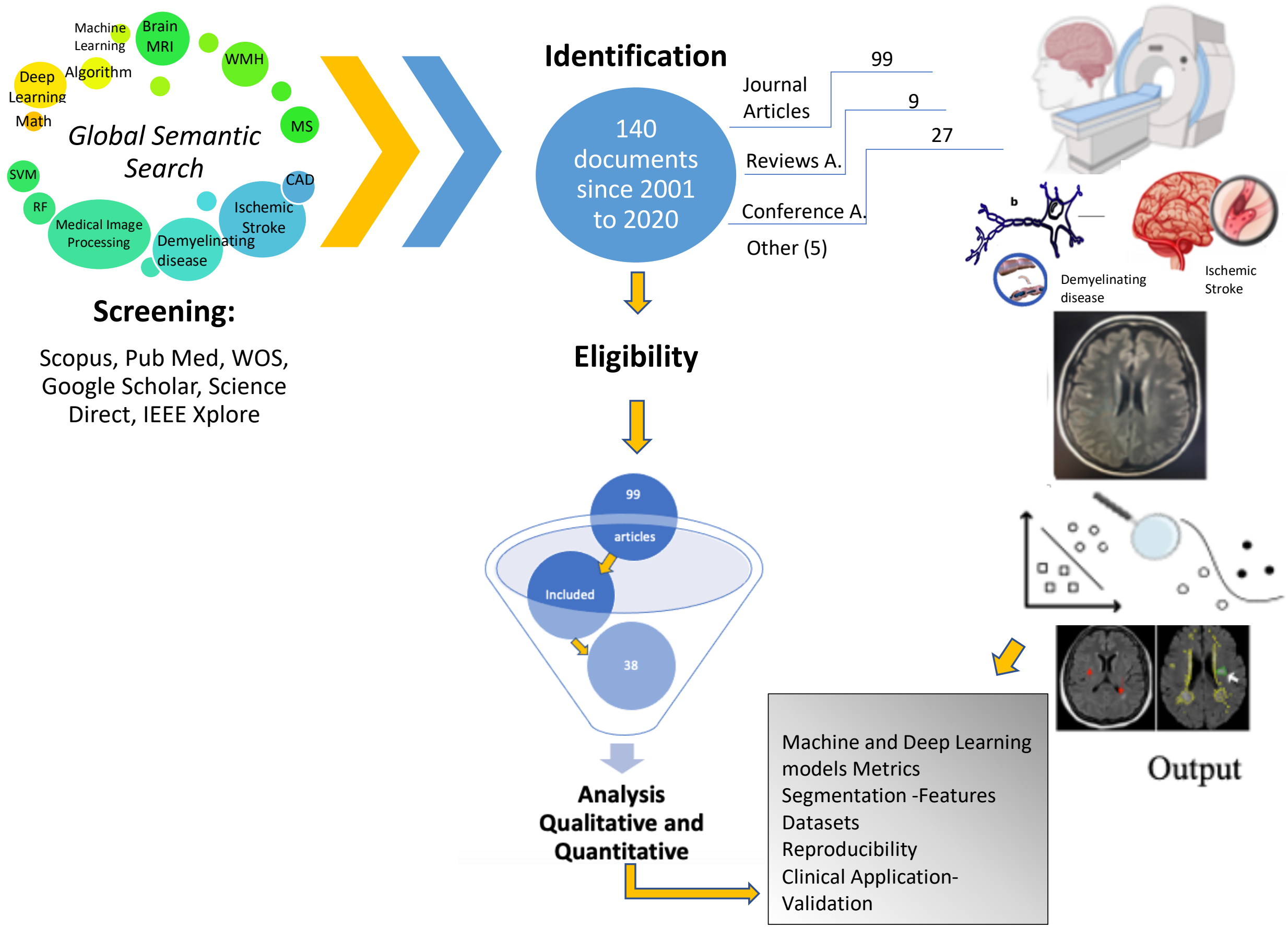

# Highlights

• New models of deep learning allow highest accuracy in detection of brain lesions

• A detailed featurization of WMH lesions is achieved combining CNN with ML models

• Deep learning models trained with small datasets have lack of reproducibility

• The CAD systems not have the enough reliability in the daily clinical application

• Bibliometrics networks can help to organize and keep a focus search of information

# 1. Introduction

There are estimated to be as many as a billion people worldwide [1] affected by peripheral and central neurological disorders [1,2]. Some of these disorders include: brain tumors, Parkinson's disease (PD), Alzheimer's disease (AD), multiple sclerosis (MS), epilepsy, dementia, neuroinfectious, stroke, and traumatic brain injuries [1]. According to the World Health Organization (WHO): ischemic stroke and "Alzheimer disease with other dementias" are the second and fifth major causes of death, respectively[2].

Biomedical images give fundamental information necessary for the diagnosis, prognosis, and treatment of different pathologies. Of all the various imaging modalities neuroimaging, is a crucial tool for studying the brain [3–9]. In terms of neuroimaging of both normal and pathologies there are different modalities namely, (1): Computed Tomography (CT), and Magnetic Resonance images (MRI) which are commonly used for the structural visualization; (2) the Positron emission tomography (PET) used principally for physiological analysis; and (3) single-photon emission tomography (SPECT) and functional MRI which are used for functional analysis of the brain [10]. Brain MRI image analysis is useful for different tasks, e.g.: lesion detection, lesion segmentation, tissue segmentation, as well as brain parcellation on neonatal, infant, and adult subjects [3,11]. MRI is frequently used in the visual inspection of cranial nerves, abnormalities of the posterior fossa and spinal cord [12], since it is less susceptible to artifacts in the image when compared to CT.

Here, we discuss the role of the MRI in the context of brain pathology, especially in the detection of White Matter Hyperintensities (WMHs). According to Leite et al., [13] due the lack of pathological studies, the etiology of WMHs are frequently proposed to be of an ischemic or demyelinating nature. Ghafoorian et al., [14] state that WMH are seen in MRI studies of neurological disorders like multiple sclerosis, dementia, stroke, cerebral small vessel disease (SVD) and Parkinson disease [13,14].

If there is an obstruction of a blood vessel , the WMH is termed an ischemia and WMH is considered to be demyelinating when there is an inflammation that causes destruction and loss of the myelin layer and compromises neural transmission [13,15,16] and this type of WMH is often related with multiple sclerosis (MS)[6,16] (**Fig 1**).

A stroke occurs when the blood flow to an area of the brain is interrupted [15,17]. There are three types of ischemic stroke according to the Banford clinical classification system [18]: (1)Partial anterior circulation syndrome (PACS) where the middle/anterior cerebral regions are affected, (2) Lacunar syndrome (LACS) where the occlusion is present in the vessels that provide blood to the deep brain regions; and (3)Total anterior circulation stroke (TACS) when middle/anterior cerebral regions are affected due to a massive brain stroke [18,19]. Ischemic stroke is a common cerebrovascular disease [1,20,21] and one of the principal causes of death and disability in low and middle-income countries[1,4,5,11,21–23]. In developed countries, brain ischemia is responsible for 75–80% of the strokes, and 10-15% are attributed to hemorrhagic brain stroke [11,19].

A demyelination disease is described as a loss of myelin with relative preservation of axons[6,16,23]. Love [16] notes that there are demyelinating diseases in which axonal degeneration occurs first and the degradation of myelin is secondary [5,16]. Accurate diagnosis classifies the demyelinating diseases of the central nervous systems (CNS) according to the pathogenesis in: "demyelination due to inflammatory processes, demyelination caused by developed metabolic disorders, viral demyelination, hypoxic-ischemic forms of demyelination and demyelination produced by focal compression"[16].

The inflammatory demyelination of the CNS is the principal cause of the common neurological disorder, Multiple Sclerosis (MS) [6,13,14,16,24]. That affect the central nervous system[25] and is characterized by lesions produced in the white matter (WM) [26] of the brain and affects nearly 2.5 million people worldwide especially young adults (ages 18-35 years) [11,24,25].

The detection, identification, classification, and diagnosis of the stroke is often based on clinical decisions made using computed tomography (CT) and MRI [27] . Using MRI, it is possible to detect the presence of little infarcts and assess the presence of a stroke lesion in the superficial and deep regions of the brain with more accuracy. This is because the area of stroke region is small and is clearly visible in MRI images compared to CT [11,15,19,20,22,28]. The delimitation of the area plays a fundamental role in the diagnosis since it is possible to misdiagnose stroke with other disorders [29,30], e.g. glioma lesions and the demyelinating diseases [13,14].

For identifying neurological disorders like stroke and demyelinating disease the manual segmentation and delineation of anomalous brain tissue is the gold standard for lesion identification. However, this method is very time-consuming, and specialist experience-dependent [19,31] and because of these limitations, automatic detection of neurological disorders is necessary even though it is a complex task because of data variability, e.g. in the case of the ischemic stroke lesions these could include the lesion shape and location, factors like the symptom onset, occlusion site, and patient differences [32].

In the past few years there has been considerable research in the field of machine learning (ML) and deep learning (DL) to create automatic or semiautomatic systems, algorithms, and methods allow detection of lesions in the brain, such as tumors, MS, stroke, glioma, AD, etc. [4,6–8,11,20,22,24,30,33–42]. Different studies demonstrate that deep learning algorithms can be successfully used for medical image retrieval, segmentation, computer aided diagnosis, disease detection and classification[43–45]. However, there is much work to be done to achieve accurate methods to get results comparable to that of specialists [37].

This critical review summarizes the literature, on deep learning and machine learning techniques in the processing, segmentation, and detection of features of WMHs found in ischemic and demyelinating dieses in brain MRI images.

The principal research questions asked here are:

- Why is research on the algorithms to identify ischemia and demyelination through the processing of medical images important?
- What are the techniques and methods used in the developing automatic algorithms for detection ischemia and demyelinating diseases in the brain?
- What are the performance metrics, and common problems of deep learning systems proposed to date?

This paper is organized as follows. Section 2 gives an outline of the selection criteria adopted for the literature review. Section 3 describes the principal machine learning and deep learning methods used in this application, and Section 4, summarizes the principal constraints, and common problems encountered in these CAD systems and we conclude this section 5 with a brief discussion.

## 2. Selection Criteria

The literature review was conducted using the recommendations given by Khan et al., [46] as well as the methodology proposed by Torres-Carrión [47,48]. We generated and analyzed bibliometric maps and identified clusters and their reference networks [49,50]. We also used the methods given in [51,52] to identify the strength of the research, as well as authors and principal research centers that work in the MRI images which use machine and deep learning for the identification of brain diseases.

The bibliometric analysis was performed by searching for the relevant literature using the following bibliographic databases[24]: Scopus[53], PubMed [54], Web of Science (WOS)[55], Science Direct[56], IEEE Xplore[57], and Google Scholar[58].

In order to perform an appropriate search it is important to focus our attention on the real theoretical context of the research, for which reason the method proposed by Torres-Carrión [48] a so-called *"conceptual mindfact"* (mentefacto conceptual), can help to organize the scientific thesaurus of the research theme [47]. **Figure 2** describes the *conceptual mindfact* used in this work to focus and constraint the topic to *MRI Brain Algorithm difference Ischemic and Demyelinating diseases* and obtain an adequate semantic search structure of the literature in the relevant scientific databases.

**Table 1** presents the semantic search structure [48] such as the input of the search specific literature (documents) in the scientific databases. The first layer is an abstraction of the *conceptual mindfact*; the second corresponds to the specific technicality namely, *Brain Processing*; the third level is relevant to the application, namely, the ischemic and demyelinating diseases. The fourth level is the global semantic structure search.

The result of the global semantic structure search (**Fig 2**) gave 140 documents related to the central theme of this work. The **Fig 3** shows the evolution of the number of publications and the type (article, conference paper and review) of the 140 documents from 2001 to December 2020. The first article related to the area of the study was published in 2001 and there has been a significant increase in the number of publications in the last three years 2018 (21), 2019 (30) and in 2020 (until December 01; 33).

**Figure 3** also shows that the journal articles predominate (99), followed by conference papers (27) and finally review articles (9). The first reviews were published in 2012(2), followed by 2013 (1), 2014 (1), 2015 (1) and in 2020 (4). Other five documents published correspond with conference review (3), editorial (1) and book chapter (1). **Fig 4** presents a list of the ten authors. Dr. Ona Wu [9,59,60] from Harvard Medical School, Boston, United States has published more documents (7) related to the research area of this review, and this correlates with his publication record as documented in the Scopus database related to ischemic stroke.

In order to analyze and answer the three central research questions of this work, the global search of the 140 documents was further refined. This filter of research complied with the categories given by Fourcade and Khonsari [61], and were applied only to the "article" documents. These criteria were;

- aim of the study: ischemia and demyelinating processing MRI brain images, identification, detection, classification or differentiate between them.
- methods: algorithms of machine learning, deep learning, neural network architectures, dataset, training, validations, testing.
- results: performance metrics, accuracy, sensibility, specificity, dice coefficient.
- conclusions: challenges, open problems, recommendations, future.

According to the second selection criteria, we found 38 documents to include in the analysis of this work and also were related to and in agreement with the items described above.

For analysis we used VOSviewer version 1.6.15 software [50] order to construct and display bibliometric maps. The data used for this objective was searched in Scopus due to its coverage of a wider range of journals [49,62].

In terms of citations and the countries of origin of these publications (**Fig 5**), we observe that the United States has a large number of citations, followed by Germany, India and the United Kingdom. This relationship was determined by the analysis of the number of citations in the documents generated by country, in agreement with the affiliation of the (primary authors? Corresponding authors?)authors and for each country the total strength of the citation link [51]. The minimum number of documents of any individual country was five and the minimum number of citations a country received was one.

The **Fig 6,** shows the network of documents and the citations, and this map relates the different connections between the documents through the citations. The scale of the colors (purple to yellow) indicate the number of citations received per document together with the year of publication and, the diameter of the points shows the normalization of the citations according to Van Eck and Waltman[52,63]. The purple points are the documents that have fewer than ten citations and yellow represents documents with more than 60 citations.

In **Table 2,** we list the ten most cited articles according to the normalization of the citations [63]. Waltman et al [51] manifest that "the normalization corrects for the fact that older documents have had more time to receive citations than more recent documents"[51,64]. Also, table two show the dataset, methodology, techniques and metrics used to develop and validate the algorithm or CAD systems proposed by theses authors.

In the bibliometric networks or science mapping there are large differences between nodes in the number of edges they have to other nodes [50]. In order to reduce these differences, the VOSviewer uses the association strength normalization [63], that is a probabilistic measure of the co-occurrence data.

The association strength normalization is discussed by Van Eck and Waltman [63], and here we construct a normalized network [50] in which the weight of the edge between nodes *i* and *j* is given by:

$s_{ij} = \frac{2ma_{ij}}{k_i k_j}$,

where $s_{ij}$ is also known as the similarity of nodes *i* and *j*, $k_i$ ($k_j$) denotes the total weight of all edges of node *i* (node *j*) and *m* denotes the total weight of all edges in the network [50].

$k_i = \sum_j a_{ij}$ and $m = \frac{1}{2} \sum_i k_i$

For more information related with normalization, mapping and clustering techniques used by VOSviewer, the reader is referred to the relevant literature [50,63,64].

From **Table 2** it can be seen that articles that are cited often deal with ischemic stroke rather than demyelinating disease. According with the methods and techniques used were support vector machine (SVM)[65], random forest (RF)[32]; classical algorithms of segmentation like Watershed algorithm (WS)[66]; and techniques of deep learning such as convolutional neural networks (CNN)[36,67]; as well as a combination between them: SVM-RF[22], CNN-RF[20,68].

# 3. Machine Learning/Deep learning methods for Ischemic Stroke and Demyelinating Disease

In the following subsections, we discuss how artificial intelligence (AI) through ML and DL methods are used in the development of algorithms for brain disease diagnosis and their relation to the central theme of this review.

## 3.1 Machine learning and deep learning:

The definitions of machine learning and deep learning are part of the global field of the Artificial Intelligence (AI) which is defined as the ability for a computer to imitate the cognitive abilities of a human being [61]. There are two different general concepts of AI: (1) *Cognitivism* related with development of rule-based programs referred to as expert systems, and (2) *Connectionism* associated to the development of simple programs educated or trained by data[61,69]. **Fig 7** presents a very general timeline of the evolution of the AI and the principal relevant facts related to the field of medicine. In addition, all applications of AI to medicine and health are not covered, e.g., ophthalmology where AI has had tremendous success, see [70–75].

Machine Learning (ML) can be considered as a subfield of artificial intelligence (AI). Lundervold and Lundervold [76] and Noguerol et al., [77] state that the main aim of ML is to develop mathematical models and computational algorithms with the ability to solve problems by learning from experiences without or with the minimum possible human intervention, in other words the model created will be able to be trained to produce useful outputs when fed input data [77]. Lakhani et al., [78] state that recent studies demonstrate that machine learning algorithms give accurate results for the determination of study protocols for both brain and body MRIs.

Machine learning can be classified into (1) supervised learning methods (e.g. support vector machine, decision tree, logistic regression, linear regression, naive Bayes and random forest), and (2) unsupervised learning methods (K-means, mean shift, affinity propagation, hierarchical clustering, and Gaussian mixture modeling) [79] (**Fig. 8)**.

*Support Vector Machine (SVM):* Algorithm used for tasks of classify, regression and clustering. SVM is driven by a linear function $w^T x + b$ similar to logistic regression [80], but with the difference that SVM only outputs class identities and does not provide probabilities. SVM classifies between two classes by constructing a hyperplane in high-dimensional feature space [81]. The class identities are positive or negative when $w^T x + b$ is positive or negative, respectively. For the optimal separating hyperplane between classes, the SVM uses different kernels (dot products) [82,83]. More information and detail about the SVM is given in the literature [80–83].

*k-Nearest Neighbors (k-NN):* The *k-NN* is a non-parametric algorithm (it means no assumption for underlying data distribution) and can be used for classification or regression[80,84]. Like a classifier k-NN is based on the measure of the Euclidean distance (distance function) and a voting function in k nearest neighbors[85] given N training vectors. The value of the *k* (the number of nearest neighbors) decides the classification of the points between classes. KNN has the following basic steps: (1) Calculate distance, (2) Find closest neighbors and (3) Vote for labels[84]. More details of the k-NN algorithm can be found in references [80,85,86]. Programming libraries such a Scikit-Learn have algorithms for k_NN [84]. The k-NN has higher accuracy and stability for MRI data, but is relatively slow in terms of computational time[86]. As an aside it is interesting to note that the nearest neighbor formulation might have been first described by the Islamic polymath Ibn al Haytham in his famous book Kitab al Manazir over a 1000 years ago ("The Book of Optics ", see: [87])!

*Random Forest (RF):* This technique is a collection of Classification and Regression Trees[88]. Here a forest of classification trees is generated where each tree is grown on a bootstrap sample of the data[89]. In that way, the RF classifier consists of a collection of binary classifiers where each decision tree casts a unit vote for the most popular class label (see **figure 8 (d)** )[90]. More information are given elsewhere [91].

*k-Means Clustering (k-means):* The k-means clustering algorithm is used for segmentation in medical imaging due to its relatively low computational complexity [92,93] and minimum computation time[94]. It is an unsupervised algorithm based on the concept of clustering. Clustering is a technique of grouping pixels of an image according to their intensity values[95,96], It divides the training set into k different clusters of examples that are near each other [80]. The properties of the clustering are measure such as the average Euclidean distance from a cluster centroid to the members of the cluster [80]. The input data for use with this algorithm should be numeric values, with continuous being better than discrete values, and the algorithm performs well when used with unlabeled datasets.

***Deep Learning methods***

Deep Learning (DL) is a subfield of ML[97] that uses artificial neural networks (ANN) to develop decision making algorithms [77]. Artificial Neural Networks are neural networks which employ learning algorithm [98] and infer rules for learning, In order to do so a set of training data examples are needed, the concept is derived from the concept of the biological neuron concept (**figure 8 (e)**). An artificial neuron receives inputs from other neurons, integrates the inputs with weights, and activates (or "fires" in the language of biology) when a pre-defined condition is satisfied [79]. There are many books describing AANs -see for example [80].

The fundamental unit of a neural network is the neuron, which has a bias w0 and a weight vector w = ($w_1$, . . ., $w_n$) as parameters θ = ($w_0$,...,$w_n$) to model a decision: $f(x) = h(w^T x + w_0)$ using a non-linear activation function h(x)[99]. The activation functions commonly used are: $\text{sign}\,(x)$ function, the sigmoid function $\sigma(x) = \frac{1}{1+e^{-x}}$ and $\tanh(x) = \frac{e^{x}-e^{-x}}{e^{x}+e^{-x}}$

An interconnected group of nodes comprise the ANN, where each node representing a neuron arranged in layers[76], the arrow representing a connection from the output of one neuron to the input of another [90]. ANNs have input layer which receives observed values, while the output layer represents the target (a value or class) and the layers between input and output layers are called hidden layers[79].

There are different types of ANNs [100] and the most common types are: convolutional neural nets (CNNs)[101], recurrent neural nets (RNN)[102], long short-term memory(LSTM)[103], and generative adversarial networks (GANs)[104]. In practice, these types of networks can be combined [100] between them and with classical machine learning algorithms.. The CNNs are most commonly used for the processing of medical images because of their success in processing and recognition of patterns in vision systems [43].

CNNs are inspired by the biological visual cortex and also called multi-layer perceptrons (MLPs)[43,105,106]. It consists of a stack of layers: convolutional, max pooling and fully connected layers. The intermediate layer is fed by the output of the previous layer e.g. the convolutional layer creates a feature map of different size and the pooling layers reduce the size of feature maps to be feed to the following layers. The final fully connected layers produce the specified class prediction at the output [43]. The general CNN architecture is presented in **Fig 9**. There is a compromise between the numbers of neurons in each layer, the connection between them and the number of layers with the number of parameters that defines the network [43]. T**able 5** presents a summary of the principal structures of CNN and the commonly used DL libraries.

More specific technical details of ML and DL, are discussed widely in the literature [7,21,42,76,76,80,98,105,107–113]. For Deep learning applications to medical image and the different architectures of neural networks and technical details, the reader is referred to various books such as Hesamian et al., [114], Goodfellow et al., [80] Zhou et al., [115], Le at al., [116], Shen et al., [105].

## 3.2 Computer-Aided Diagnosis in Medical Imaging (CADx system)

Computer-aided diagnosis has its origins in 1980s at the Kurt Rossmann Laboratories for Radiologic Image Research in the Department of Radiology at the University of Chicago, [117]. The initial work was on detection of breast cancer [29,117,118] and the reader is referred to a recent review [119].

There has been much research and development of CADx systems using different modalities of medical images. The CAD not is a substitute for the specialist but can assist or be an adjunct to the specialist in the interpretation of the images [34]. In other words, CADx systems can provide a "second objective opinion" [89,99] and make the final disease decision from image-based information and the discrimination of lesions, complementing a radiologist's assessment[120].

CAD development takes into consideration the principles of radiomics [40,121–125]. The term radiomics is defined as the extraction and analysis of quantitative features of medical images, in other words the conversion of medical images into mineable data with high fidelity and high throughput for decision support [40,121,122]. The medical images used in radiomics are obtained principally with CT, PET or MRI [40].

The steps that are utilized by a CAD system consists of [40]: (a) image data and preprocessing, (b) image segmentation, (c) feature extraction and qualification, (d) classification (**Fig 10).**

### 3.1 Image Data:

The dataset is the principal component to develop an algorithm because it is the nucleus of the processing. Razzak et al., [109] state that the accuracy of diagnoses of the disease depends upon image acquisition and image interpretation. However Shen et al., [105] add a caveat that the image features obtained from one method need not be guaranteed for other images acquired using different equipment [105,126,127]. For example, it has been shown that the methods of image segmentation and registration designed for 1.5-Tesla T1-weighted brain MR images are not applicable to 7.0-Tesla T1-weighted MR images [43,57,58].

There are different datasets of images for brain medical image processing,  in the case of stroke, the most famous datasets used is the ISLES (Ischemic Stroke Lesion Segmentation) dataset [20,68] and ATLAS (Anatomical Tracings of Lesions After Stroke) [128]; for the case of demyelinating disease there isn't a specific dataset, but datasets for Multiple Sclerosis is used, e.g., MSSEG (MS segmentation)[129]. **Table 3** describes the datasets that have been used in the publications under consideration in this review is possible find datasets for brain medical image processing.

### 3.2.1 Image Preprocessing:

There are several preprocessing steps necessary to reduce the noise and artifacts in the medical images, before the segmentation[34,130,131].

The preprocessing  steps commonly used are (1) the *grayscale conversion*, and *the image resizing* [131]   to get better contrast and enhancement, (2)  *bias field correction* to correct the intensity inhomogeneity [24,130], (3) *image registration*, a process for spatial alignment [130], and  (4) *removal of nonbrain tissue*  such as fat, skull, or neck  which have intensities overlapping with intensities of brain tissues [21,130,132].

### 3.2.2 Image Segmentation:

In simple terms image segmentation is the procedure of separating a digital image into a different set of pixels[31] and is considered the most fundamental process as it extracts the region of interest (ROI) through a semiautomatic or automatic process[133]. It divides the image into areas according a specific description to obtain the anatomical structures and patterns of diseases.

Despotovíc et al., [130], Merjulah and Chandra [31] indicate that the principal goal of the medical image segmentation is to make things simpler and transform it *into a set of semantically meaningful, homogeneous, and nonoverlapping regions of similar attributes such as intensity, depth, color, or texture* [130] because the segmentation assists doctors to diagnose and make decisions[31].

According to Despotovíc et al.,[130] the segmentation methods for brain MRI are classified into: (i) manual segmentation, (ii) intensity-based methods (including thresholding, region growing, classification, and clustering), (iii) atlas-based methods, (iv) surface-based methods (including active contours and surfaces, and multiphase active contours), and (v) hybrid segmentation methods[130].

To evaluate, validate and measure the performance of every automated lesion segmentation methodology compared to the expert segmentation [134] one needs to consider the accuracy (evaluation measurements) and reproducibility of the model [135]. The evaluation measurements compare the output of segmentation algorithms with ground truth in either a pixel-wise or a volume-wise basis [3].

The accuracy is related with the grade of closeness of the estimated measure to the true measure [135], and for that are possible four situations: true positives (TPs) and true negatives (TNs), where in the segmentation is correct, and false positives (FPs) and false negatives (FNs), where in there is disagreement between the two segmentations.

the most commonly used metrics for evaluate the automatic segmentation accuracy, quality and the strength of the model are [136]:

- *Dice similarity coefficient (DSC):* gives a measure of overlap between two segmentations (computed and corresponding reference) and is sensitive to the lesion size. A DSC of 0 indicates no overlap, and a DSC of 1 indicates a perfect overlap, 0.7 normally is considered a good segmentation [32,37,135–138].

$$DSC = \frac{2TP}{FP+FN+2TP}$$

- *Precision:* is the measure of over-segmentation between 0 and 1, and it means the proportion of the computed segmentation which overlaps with the reference segmentation[136,137], This is also is called the positive predictive value (PPV), with a high PPV indicating that a patient identified with a lesion does actually have the lesion[139].

$$Precision = \frac{TP}{FP+TP}$$

- *Recall also known as Sensitivity:* Gives a metric between 0 and 1, this a sign of over-segmentation, and it is a measure of the amount of the reference segmentation which overlaps with the computed segmentation [136,137].

  $$Recall = Sensitivity = \frac{TN}{TN + FN}$$

  The metrics of overlap measures which are less often used are the sensitivity, specificity (measures the portion of negative voxels in the ground truth segmentation[140]) and accuracy, which according with García-Lorenzo et al.,[135] and Taha and Hanbury[140], should be considered carefully because they penalize errors in small segments more than in large segments. These are defined as:

  $$Specificity = \frac{TN}{FP + TN}$$

  $$Accuracy = \frac{TP + TN}{TP + FP + FN + TN}$$

- *Average Symmetric Surface Distance (ASSD, mm):* represents the average surface distance between two segmentations (computed and reference and vice-versa), and is an indicator of how well the boundaries of the two segmentations align. ASSD is measured in millimeters, and a smaller value indicates higher accuracy[68,134,137].
  The average surface distance (ASD) is given as:

  $$ASD(X,Y) = \sum_{x \in X} min_{y \in Y} d(x,y)/|X|$$

  Where $d(x,y)$ is a 3D matrix consisting of the Euclidean distances between the two image volumes $X$ and $Y$, and $ASSD$ is defined[134]:

  $$ASSD(X,Y) = \{ASD(X,Y) + ASD(Y,X)\}/2$$

- *Hausdorff's distance (HD,mm):* It is more sensitive to segmentation errors appearing away from segmentation frontiers than ASSD[137]. The Hausdorff measure is an indicator of the maximal distance between the surfaces of two image volumes (the computed and reference segmentations)[20,137]. HD is measured in millimeters and like the ASSD a smaller value indicates higher accuracy[134].

$$d_H(X,Y) = max\{max_{x \in X} min_{y \in Y} d(x,y), max_{y \in Y} min_{x \in X} d(y,x)\}$$

where $x$ and $y$ are points of lesion segmentations $X$ and $Y$, respectively, and $d(x,y)$ is a 3D matrix consisting of all Euclidean distances between theses points[134].

- Intra Class Correlation (ICC): Is a measure of correlation between volumes segmented and ground truth lesion volume [137].

- Correlation with Fazekas score: A Fazekas score is a clinical measure of WMH, comprising of two integers in the range [0, 3] reflecting the degree of periventricular WMH and deep WMH respectively [137].

- *Relative Volume difference (VD, %):* It measure the agreement between lesion volume and the ground truth lesion volume, a low VD means more agreement [139,141].

$$VD = \frac{(v_s - v_g)}{v_g}$$

where, $v_s$ and $v_g$ are the segmented and ground truth lesion volumes respectively.

Lastly, we define [135] the "reproducibility" which is a measure of the degree of agreement between several identical experiments. Reproducibility guarantees that differences in segmentations as a function of time result from changes in the pathology and not from the variability of the automatic method [135].

**Tables 2 and 4** presents the types of databases, modalities and the evaluation measurements considered and applied to the results reported in the literature to date.

### 3.2.3 Feature extraction:

An ML or DL algorithm is often a classifier [113] of objects (e.g. lesions in medical images). Feature selection is a fundamental step in the processing of medical image and more specially it allows us to research which features are relevant for the specific classification problem of interest, and also it helps to get higher accuracy rates [42].

The task of feature extraction is complex due to the task of determining to determine an algorithm that can extract a distinctive and complete feature representation, and for that principal reason it is very difficult to generalize and implies that one has to design a featurization method for every new application [99]. In DL, this process is also denoted to as "hand-crafting" features [99].

The classification is related to the extracted features that are entered as input to an ML model [113], while a DL algorithm model uses pixel values in images directly as input information instead of features calculated from segmented objects [113].

In the case of processing stroke with CNNs the featurization of the images is a key application [68,142] and depends on the signal-to-noise ratio in the image, which can be improved by target identification via segmentation to select regions of interest [142]. According to Praveen et al., [143], a CNN learns to discriminative local features and return better performance than handcrafted features.

Texture analysis is a common technique in medical pattern recognition tasks to determine the features, and for that one uses second-order statistics or co-occurrence matrix features[40]. Mitra et al., [139], indicate that they derive local features, spatial features and context-rich features from the input MRI channels.

It is clear that currently the DL algorithms especially those that use of a combination of CNNs and machine learning classifiers produce a marked transformation [144] in the featurization and the segmentation in medical image processing [76,142]. CNNs have a high utility in tasks like identification of compositional hierarchy features and low-level features (e.g. edges), specific pattern forms and intrinsic structures (e.g. shapes, textures) can be developed [3] and spatial features generated from an n-dimensional array of basically any arbitrary size[37,108]. e.g. the U-Net model proposed by Ronneberger et al., [145], employed parameter sharing between encoder-decoder paths for incorporating spatial and semantic data that allow better segmentation performance [136]. Based on the U-Net model, currently there are novel variants of U-Net designs, e.g. Bamba et al., [146], used a U-net architecture with 3D convolutions that allow the use of an attention gate for the decoder to suppress unimported parts of the input while emphasize the relevant features. There is considerable room for improvement and innovation of innovative networks (e.g., [147]).

The process of converting a raw signal into a predictor (automatization of the featurization) constitutes an advantage of the DL methods over others, which is useful when there are large volumes of data of uncertain relationship to an outcome[142], e.g. the featurization of acute stroke and the demyelinating diseases.

**3.3 ML and DL classifiers applied to diagnosis ischemia and demyelinating diseases**

In this subsection we discuss the different classifiers that have been utilized in the literature under. Additional details such as dataset and the measure metrics of the algorithms and the tasks are presented in the **Tables 2 and 4.**

In the **table 4** is noted that even though there are a large number of publications related to stroke ischemia (27 documents) and most deal with classification of stroke patients versus normal controls, or prediction of post-stroke functional impairment or treatment outcome [15,19,20,22,27,32,36,59,60,65,67,68,131,132,136,143,148–158], there is a paucity of results related to demyelinating disease alone. However there are some publications dealing with Multiple Sclerosis (MS) which is the most common demyelinating disease (2) [159,160]. In addition, there are articles related to WMHs (5) as well as articles that combine the ischemic stroke with MS and other brain injuries like gliomas (4) [30,137,164,165].

Different studies [15,65,79,158] related to stroke (see **table 4** and **fig 1**) in their different types, use principally classifiers of ML to determine the properties of the lesion. The classifiers most commonly used are SVM and Random Forest (RF)[158].

According to Lee et al., [158] the RF has some advantages over the SVM because RF can be trained quickly and provides insight into the features that can predict the target outcome[158]; also the RF can automatically perform the task of feature selection and provide a reliable feature importance estimate. Additionally, the SVM is effective only in cases where the number of samples is small compared with the number of features [79,158]. Along similar lines, Subudhi et al.,[22] reported that the RF algorithm works better when one has a large dataset and it is more robust when there are a higher number of trees in the decision making process, They reported an accuracy of 93.4% and DSC index of 0.94 in their study.

Huang et al., [65] present results that predict ischemic tissue fate pixel-by-pixel based on multi-modal MRI data of acute stroke using a flexible support vector machine algorithm [65]. Nazari-Farsani et al.,[27] proposes an identification of the ischemic stroke through SVM with Linear Kernel and cross validation folder with an accuracy of 73% with a private dataset of 192 patients scans, while Qiu et al.,[151] with a private dataset of 1000 patients for the same task use only the Random Forest (RF) classifier and obtain an accuracy of 95%.

The combination of the traditional classifier likes SVM and RF with CNN show better results, e.g. [32,65,143] report values of DSC between 0.80 and 0.86. Melingi and Vivekanand [131] reported that through combination of the Kernelized Fuzzy C-Means clustering and SVM they achieved an accuracy of 98.8% and sensitivity of 99%.

A method for detecting the stroke presence or non-stroke presence using the SVM and feed-forward backpropagation neural networks classifiers, is presented in [15]. For extraction of the features of the segmentation of the stroke region a k-means clustering was used along with adaptive neuro fuzzy inference system (ANFIS) classifier, since the other two methods failed to detect the stroke region in low edges brain images, resulting in the accuracy and the precision of 99.8% and 97.3% respectively.

The different developments of architectures in the DL models contribute to get better evaluation and results of segmentation, e.g. Kumar et al., [136] proposed a combination of U-Net and Fractal Networks, Fractal networks, are based on the repetitive generation of self-similar objects and ruling out residual connections [136,166]. They report on Sub-acute stroke lesion segmentation (SISS) and acute stroke penumbra estimation (SPES) using a public database (ISLES 2015, ISLES 2017), an accuracy of 0.9908 and DSC of 0.8993 for SPES, and values of accuracy: 0.9914 and DSC of 0.883 for SISS. Clèrigues et al., [156] with the same public database and tasks proposed the uses of U-Net based 3D CNN architecture and 32 filters and obtained values of DSC: 0.59 for SISS and for SPEC and DSC of 0.84.

Multiple sclerosis (MS) is characterized by the presence of white matter (WM) lesions and constitutes the most common inflammatory demyelinating disease of the central nervous system[6,167] and for that reason is often confused with other pathologies, since the key for that is the determination and characterization of the WMHs. Guerrero et al.,[161] using CNNs with u-shaped residual network architecture (uResNet) with the principal task of differentiating the WMHs, found values of DSC69.5 for WMH and DSC: 40.0 for ischemic stroke.

Mitra et al.[139] in their work of lesion segmentation also presented differentiation of the ischemic stroke and MS through the analysis of the WMHs and reported a DSC of 0.60, while using only the classical RF classifier. Similar work by Ghaforian et al.,[14] but with the central aim of determining the WMHs that correspond to Cerebral small vessel disease (SVD), reported a sensitivity of 0.73 with 28 false positives using a combination of AdaBoost and RF algorithms,.

# 4. Common Problems in medical image processing for ischemia and demyelinating brain diseases

This section presents a brief summary of some common problems found in the processing of ischemia and demyelinating disease images.

*4.1 The dataset*

The availability of large datasets is a major problem in medical imaging studies, and there are few datasets related to specific diseases [27]. The lack of datasets is a challenge since deep learning methods require a large amount of data for training, test and validation [27].

Another major problem is that even though algorithms for ischemic stroke segmentation in MRI scans have been (and are) intensively researched, but the reported results in general do not allow us to establish a comparative analysis due to the use of different databases (privates and public) with different validation schemes [29,34].

The Ischemic Stroke Lesion Segmentation (ISLES) challenge, was designed to facilitate the development of tools for the segmentation of stroke lesions [20,68,142]. The Ischemic Stroke Lesion Segmentation (ISLES) group [20,68] have a set of stroke images, but there is a need to enrich the dataset with clinical information, in order to get better performance with CNNs.

Another problem with the datasets, is the need for accurately labeled data [37], This lack of annotated data constitutes a major challenge for ML supervised algorithms [168] because the methods have to learn and train with limited annotated data which in most cases contain weak annotations (sparse annotations, noisy annotations, or only image level annotations) [144]. Therefore collecting image data in a structured and systematic way is imperative [79] due the large database required by the AI techniques to function efficiently.

An example of good practice of health data (images and health information) is exemplified by the UK Biobank [169], which has health data from half a million UK participants. The UK Biobank aims to create a large-scale biomedical database that can be accessed globally for public health research. However, the access depends on administrator approval and payment of a fee.

Other difficulties that accompany the labeling of the images in a dataset include the lack of collaboration between clinical specialists and academics, patient privacy issues, and the most importantly the costly time-consuming task of manual labeling of data by clinicians[34].

With CNNs overfitting is a common problem due the small size of the training data [114], and therefore it is important the increase of the size of training data and, one solution for this problem is the use of the technique of "data augmentation" which according to [170], helps improve generalization capabilities of deep neural networks, and can be perceived as implicit regularization, e.g. Tajbakhsh et al., in [144,171] reported in their results that the sensitivity in a model improves 10% (from 62 to 72%) if the dataset is increased from a quarter to full size of the training dataset. Various methods of data augmentation of medical images are reviewed in [172].

However, in [141] it is suggested that cascaded CNN architectures are a practical solution for the problem of the limited annotated data, in that the proposed architecture tends to learn well from small sets of data [141].

An additional but no less important problem, is the availability of equipment for collecting the image data. Even though the MRI is better than CT for stroke diagnosis [173] there is also the fact that in some developing countries the availability of CT and MRI facilities is very limited and relatively expensive, in addition to lack of trained technical personnel and information [34]. Even in developed countries there are disparities in availability of equipment between urban and rural areas. These issues are discussed for example in a report published by the Organization of Economic Cooperation and Development (OECD) [174].

*4.2 Detection of lesions*

It is known that that the brain lesions have a high degree of variability [8,64], e.g., stroke lesions and tumors, and hence it is a hard and complex challenge to develop a system with great fidelity and precision. As an example, the lesion size and contrast affect the performance of the segmentation [18].

In the case of the WMHs and their association with a determined disease like the ischemic stroke, demyelinating disease or any other disorders, the set of features to describe their appearances and different locations [14], plays a fundamental role for training with the minimum errors of any model.

*4.3 Computational cost*

In medical image processing the computational cost is a fundamental factor, since the ML algorithms often require a large amount of data to ‘‘learn’’ to provide useful answers [100] and hence increased computational costs. Different studies [110,113,175] report that training neural networks which are efficient and make accurate predictions have a high computational cost (e.g. time, memory, and energy) [110]. This problem is often a limitation with the CNNs due to the high dimensionality of input data and the large number of training images required [113]. However graphical processing units (GPUs) have proven to be flexible and efficient hardware for ML purpose [100]. GPUs are highly specialized processors for image processing. The area of General purpose GPU (GPGPU) Computing is a growing area and is an essential part of many scientific computing applications. The basic architecture of a GPU differs a lot from a CPU. The GPU is optimized for high computational power and high throughput. CPUs are designed for more general computing workloads. GPUs in contrast are less flexible, however GPUs are designed to compute in parallel the same instructions. As noted earlier neural networks are structured in a very uniform manner such that at each layer of the network identical artificial neurons perform the same computation. Therefore the structure of a network is highly appropriate for the kinds of computation that a GPU can efficiently perform. GPUs have other additional advantages over CPUs, such as more computational units and a higher bandwidth to retrieve from memory. Furthermore, in applications requiring image processing GPU graphics specific capabilities can be exploited to further speed up calculations. As noted by Greengard "Graphical processing units have emerged as a major powerhouse in the computing world, unleashing huge advancements in deep learning and AI" [176,177].

Suzuki et al., [113,178] propose the utilization of massive-training artificial neural network (MTANN) [179] instead of the CNNs because the CNN requires a huge number of training images (e.g., 1,000,000), while that the MTANN requires a small number of training images (e.g., 20) because of its simpler architecture. They note that with GPU implementation, an MTANN completes training in a few hours, whereas a deep CNN takes several days [113], of course, currently this depends upon the task as well as the processor speed.

It has been proposed that one can use small convolutional kernels in 3D CNNs [144]. This architecture seems to be more discriminative without increasing the computational cost and number of trainable parameters in relation to the task of identification [164].

## 5. Discussion and conclusions

The techniques of deep learning are going to play a major role in medical diagnosis in the future, and even with the high training cost, CNNs appear to have great potential and can serve as a preliminary step in the design and implementation of a CAD system[34].

However, brain lesions, and especially the WMHs have significant variants with respect to size, shape, intensity, and location, which makes their automatic and accurate segmentation challenging [159]; e.g. in spite of the fact that stroke is considered to be easy to recognize and differentiate from other WMHs for experienced neuroradiologists, it could be a challenge and difficult task for general physicians, especially in rural areas or in developing countries where there are shortages of radiologists and neurologists and, for that reason it is important to employ computer-assisted methods as well as telemedicine[180], in this sense, e.g. Mollura et al., [181] gives some strategies in order to get an effective and sustainable implementation of radiology in developing countries.

Our research has noted diverse approaches in the detection differentiation of WHMs, especially with ischemic stroke and demyelinating disease like MS. Those include methods like support vector machine (SVM), neural networks, decision trees, or linear discrimination analysis.

In the ISLES 2015 [20] and ISLES 2016 [68] competitions the best results were obtained for stroke lesion segmentation and outcome prediction using the classic machine learning models, specifically the Random Forest (RF); whereas in ISLES 2017 [68] the participants offered algorithms that use CNN, but the overall performance was not much different from ISLES 2016. However, the ISLES team state that despite this deep learning has the potential to influence clinical decision making for stroke lesion patients[68]. However, this is only in the research setting and has not been applied to a real clinical environment, in spite of development of many CAD systems [100].

To identify stroke, according to Huang et al.,[65] the SVM method provides better prediction and quantitative metrics compared with the ANN. Also, they note the SVM provides accurate prediction with a small sample size [65,182], Feng et al., [142] indicate that the biggest barriers in applying deep learning techniques to medical data are the insufficiency of the large datasets that are needed to train DNNs[142].

Although various models trained with small datasets report good results (DSC values > 0.90) in their classifications or segmentations, see table 4 [15,148,152], Davatzikos [183] recommends avoidance of methods trained with small datasets because of replicability and reproducibility issues [77,183]. Therefore, it is important to have multidisciplinary groups [77,98,184] involving representatives from the clinical, academic and industrial communities in order to create efficient processes that can validate the algorithms and hence approve or refute recommendations made by software [77]. Related to this is that algorithmic development has to take into consideration that real life performance by clinicians is different from models.

However other areas of medicine, for example ophthalmology has shown that certain classifiers approach clinician level performance. Of further importance is the development of explainable AI methods which have been applied to ophthalmology where correlations are made between areas of the image that the clinician uses to make decisions and the ones used by the algorithms to arrive at the result (i.e., the portions of the image which most heavily weighs the neural connexons)[71,185–187].

Thus, the importance of involving actively clinical AI research, multidisciplinary communities, it is possible to pass the "valley of death"[100] namely the lack of resources and expertise often encountered in translational research. This will take into account the fact that currently deep learning is a black box [43], where the inputs and outputs are known but the inner representations are not well understood. This is being alleviated by the development of explainable AI [72].

Even though there have been tremendous advances, there are only a few methods that are able to handle the vast range of radiological presentations of subtle disease states. There is a tremendous need for large annotated clinical data sets, a problem that can be (partially) solved by data augmentation and by methods of transfer learning [188,189] used in the models principally with different CNNs architectures.

Although it is very important to note that processing diseases or tasks in medical images are not the same as processing general pictures of say, dogs or cats, but it is possible uses a set of generic features already trained in CNNs for a specific task to transfer as features for input to classifiers focused on other medical imaging tasks. For examples in medical imaging see: [190–193].

Finally it is important keep in mind the fact of the mentioned by [194] that like humans, the software is only as good as the data it is trained on. Therefore, it is important that research in medical image analysis and diagnosis must include both clinical and technical knowledge.

**Acknowledgements:**

VL acknowledges the award of a DISCOVERY grant from the Natural Sciences and Engineering Research Council of Canada for research support.

D.C. acknowledges "the mobility scholarship 2020 from Universitat Politècnica de València" for research stay. D.C. also, acknowledges the research support of Universidad Técnica Particular de Loja through the project PROY_INV_QUI_2020_2784.

M.J.R.-Á. the Spanish Government Grant PID2019-107790RB-C22, "Software development for a continuous PET crystal system applied to breast cancer" for research support.

## Figure Captions

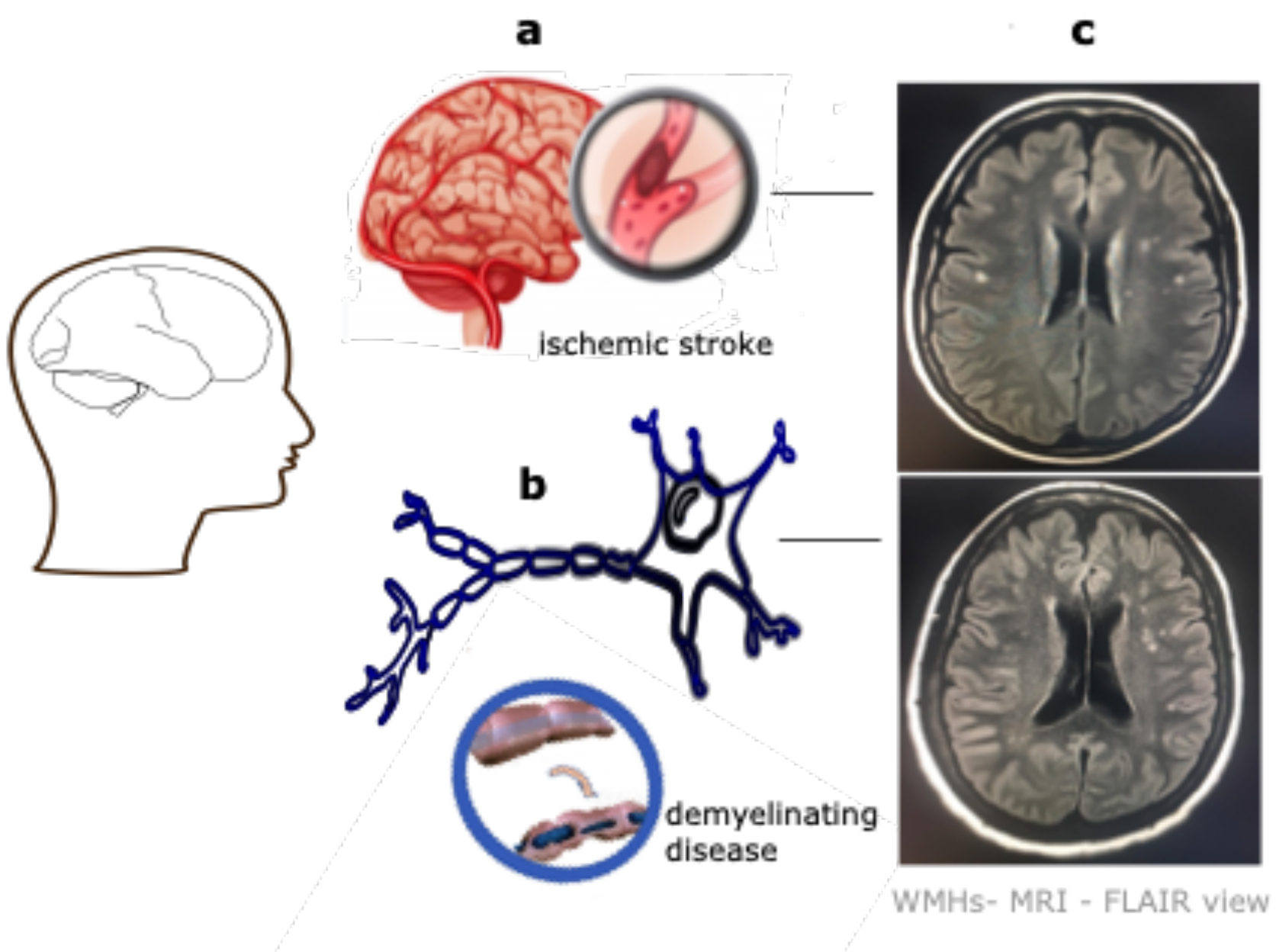

**Fig 1.** Diseases focused in this review: a) Ischemic stroke, occurred when a vessel in the brain is blocked; b) Demyelinating disease is the loss of myelin layer in the axon of neurons; c) The WMHs of the ischemic stroke and demyelinating view through MRI-FLAIR modality, without an expert and trained eye specialist, it is difficult to distinguish one disease to another by their similarities in the WMHs.

.

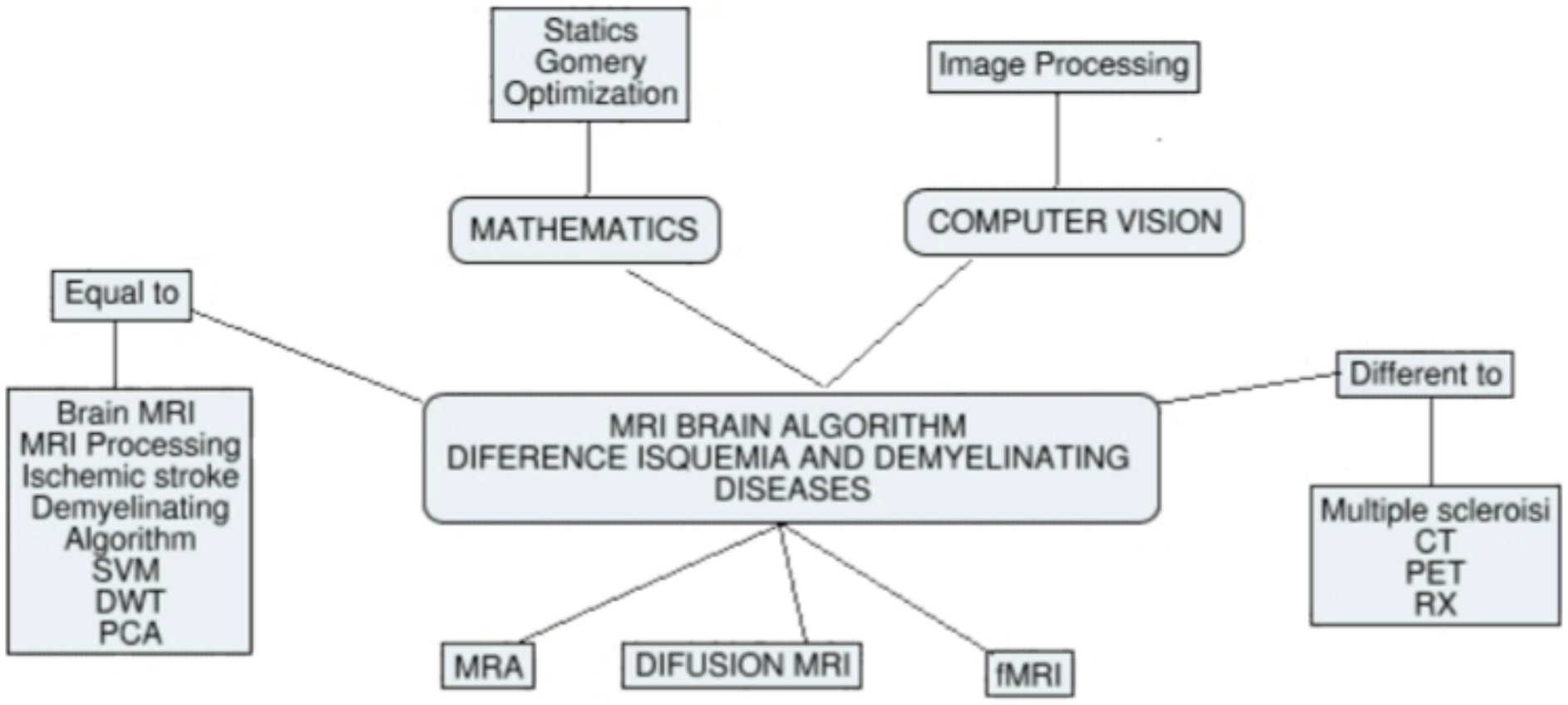

**Fig 2.** Conceptual Mindfact (Mentefacto conceptual) according to [47,48]. This allows the keyword identification for a systemic search of the literature in the scientific databases.

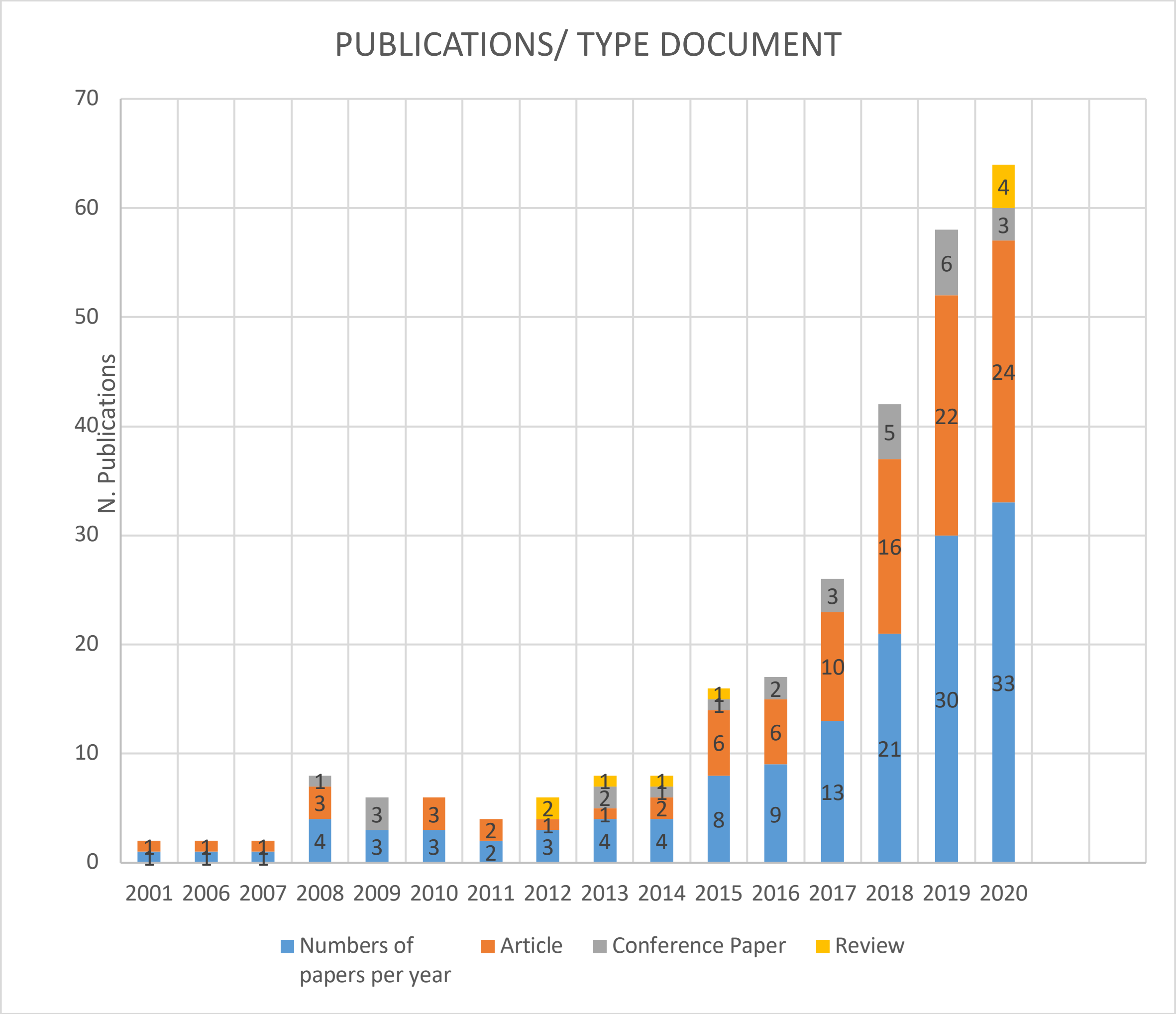

**Fig 3.** Evolution of the number of publications and the type (article, conference paper and review) of the 140 documents from 2001 to November 2020. The first article related to theme of this work was published in 2001, in the period 2002-2005 there were not published any document and the maximum number of publications is in 2020 with 33 documents. In relation to type of documents: articles are the greatest number of publications (99), followed by conference papers (27) and finally review articles (9). Other five documents published correspond with conference review (3), editorial (1) and book chapter (1). The reviews started to publish only until 2012 (2), 2013 (1), 2014 (1), 2015 (1) and in 2020 (4).

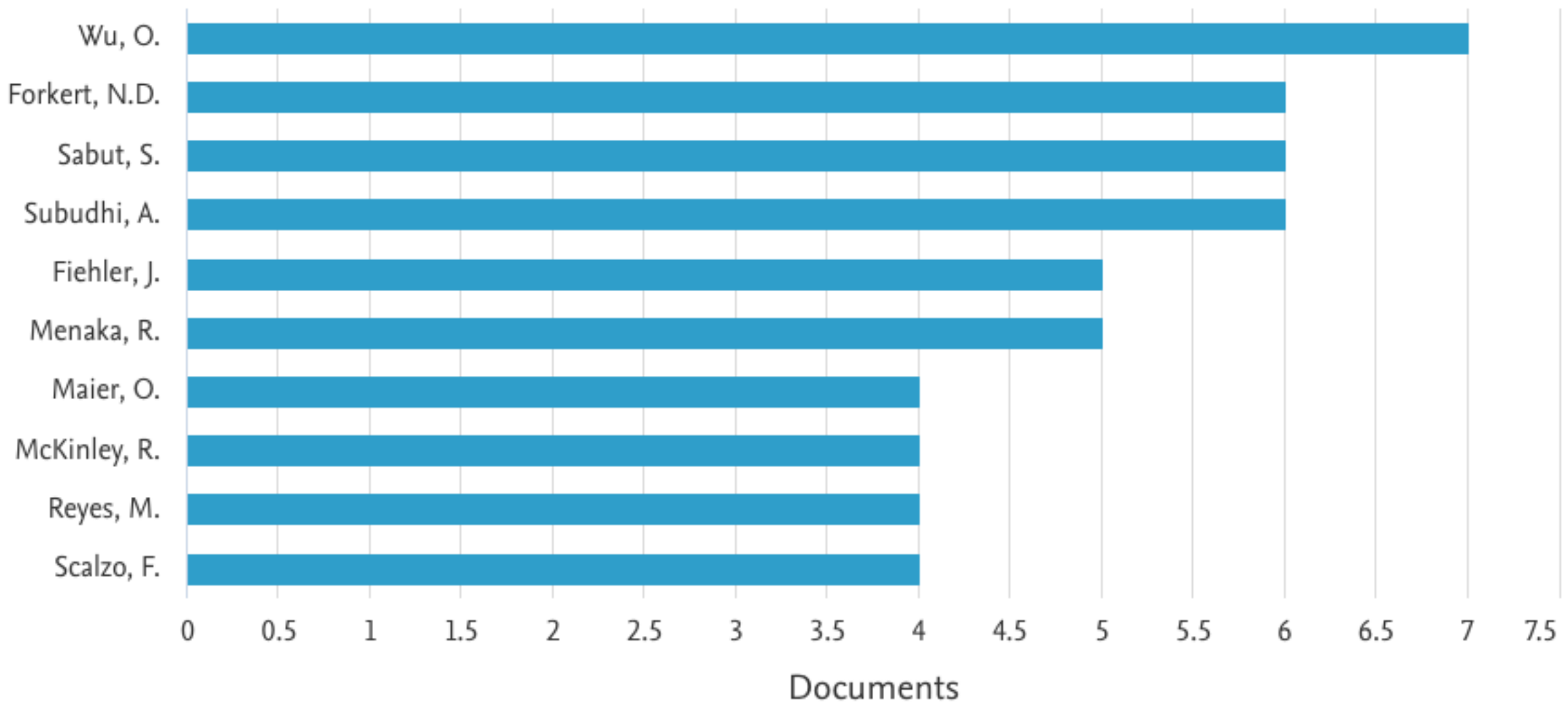

**Fig 4.** Classification of the first ten authors with more documents published according with the first criteria of search. In the figure, it can be seen that author Dr. Ona Wu [9,59,60] from Harvard Medical School, Boston, United States has more documents (7) related with the research area of this review, also this principal theme of research in agreement of his historical of publications in Scopus database is related with the ischemic stroke.

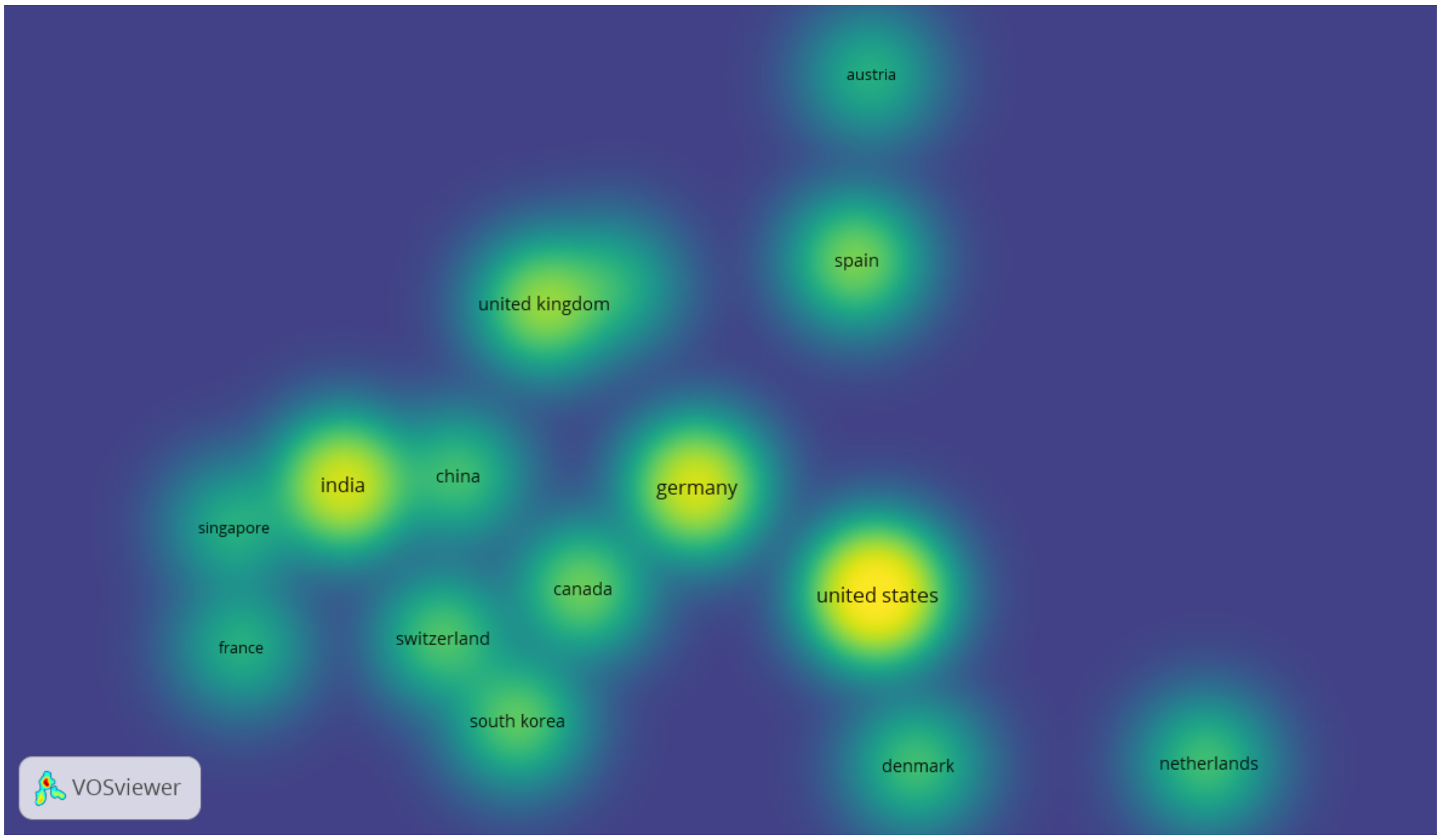

**Fig 5.** Network of the publications in relation with the citations and the countries of documents. The countries were determined by the first authors affiliation. In the map the density of yellow color in each country indicates the quantity of citations: United States has a strong number of citations in their documents, followed up Germany, India and United Kingdom.

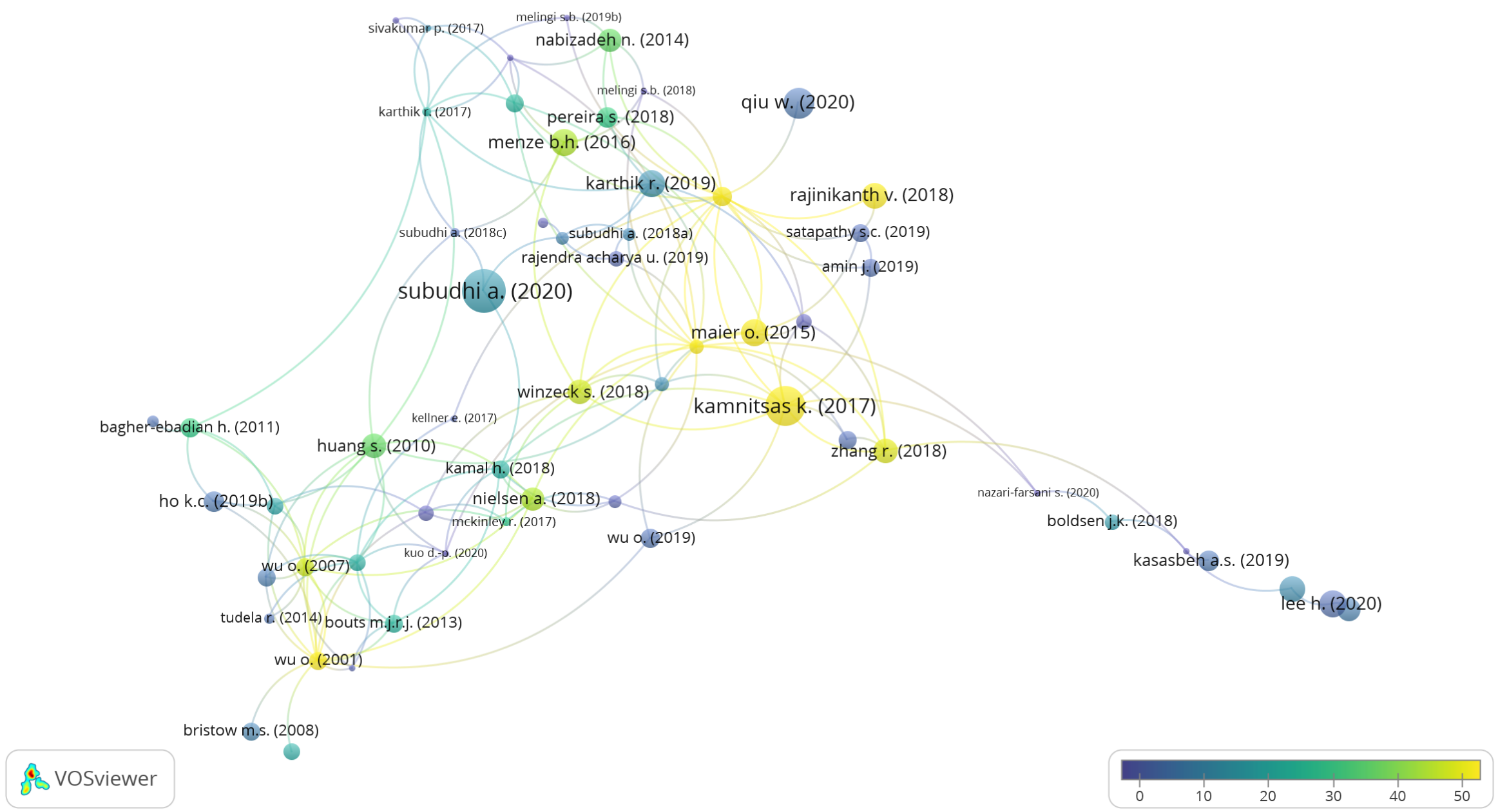

**Fig 6**. Citation map between documents generated in VOSviewer [50]. The scale of the colors (purple to yellow) indicate the number of citation per document and the diameter of the points shows the normalization of the citations according to Van Eck and Waltman[63]. The purple points are the documents that have less than ten citations and the yellow points are the documents with more than 60 citations.

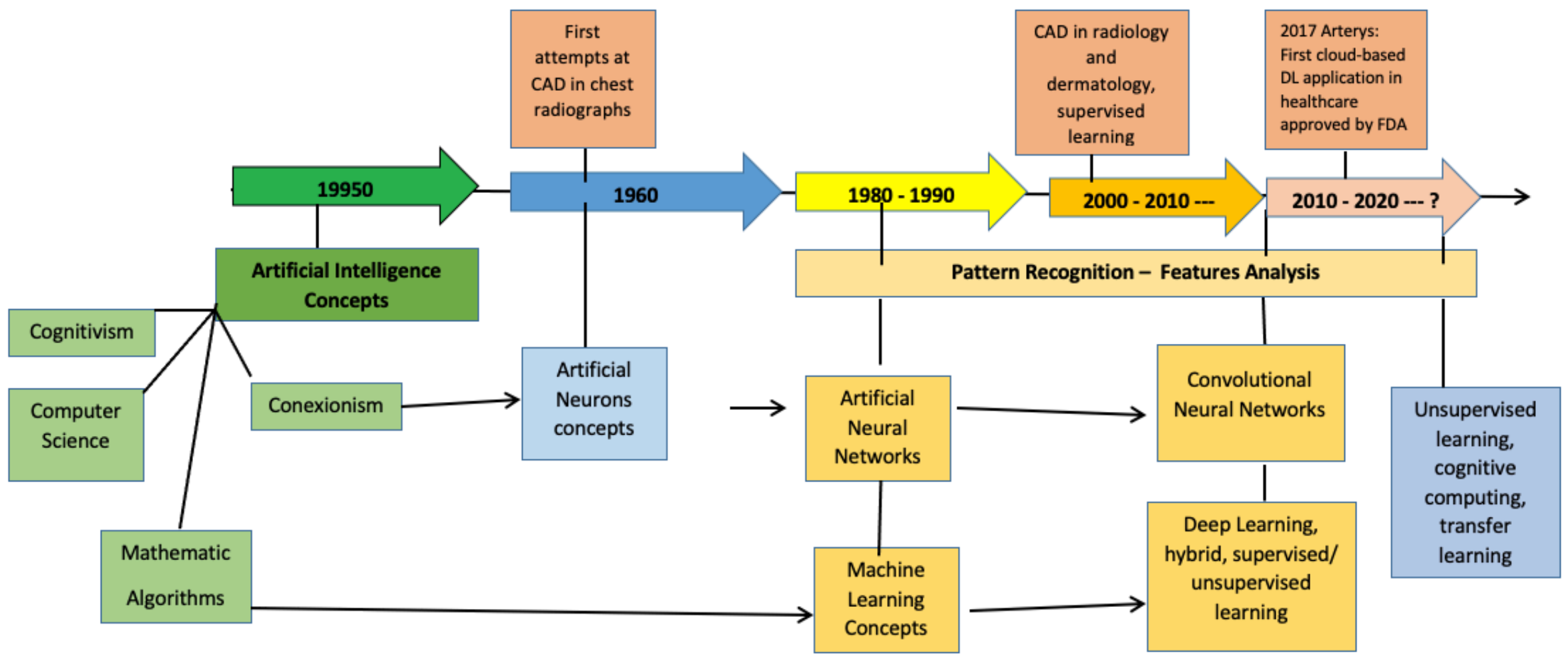

**Fig 7**. A general timeline of the evolution of the AI (lowest level) and the principal relevant applications related in the field of medicine (upper level), since 1950 to currently. It shows also the relation of the initial concepts and their derivation in the machine and deep learning, also is shown that pattern recognition is an important factor in the evolution since the born of the Artificial Neural Networks concepts in 1980 to currently where is added the analysis of the features. In the case of the applications, the Arterys model based in DL and approved by FDA in 2017 is a sample of the increasing interest of research in the field of health. This figure was created and adapted using references [61,195,196].

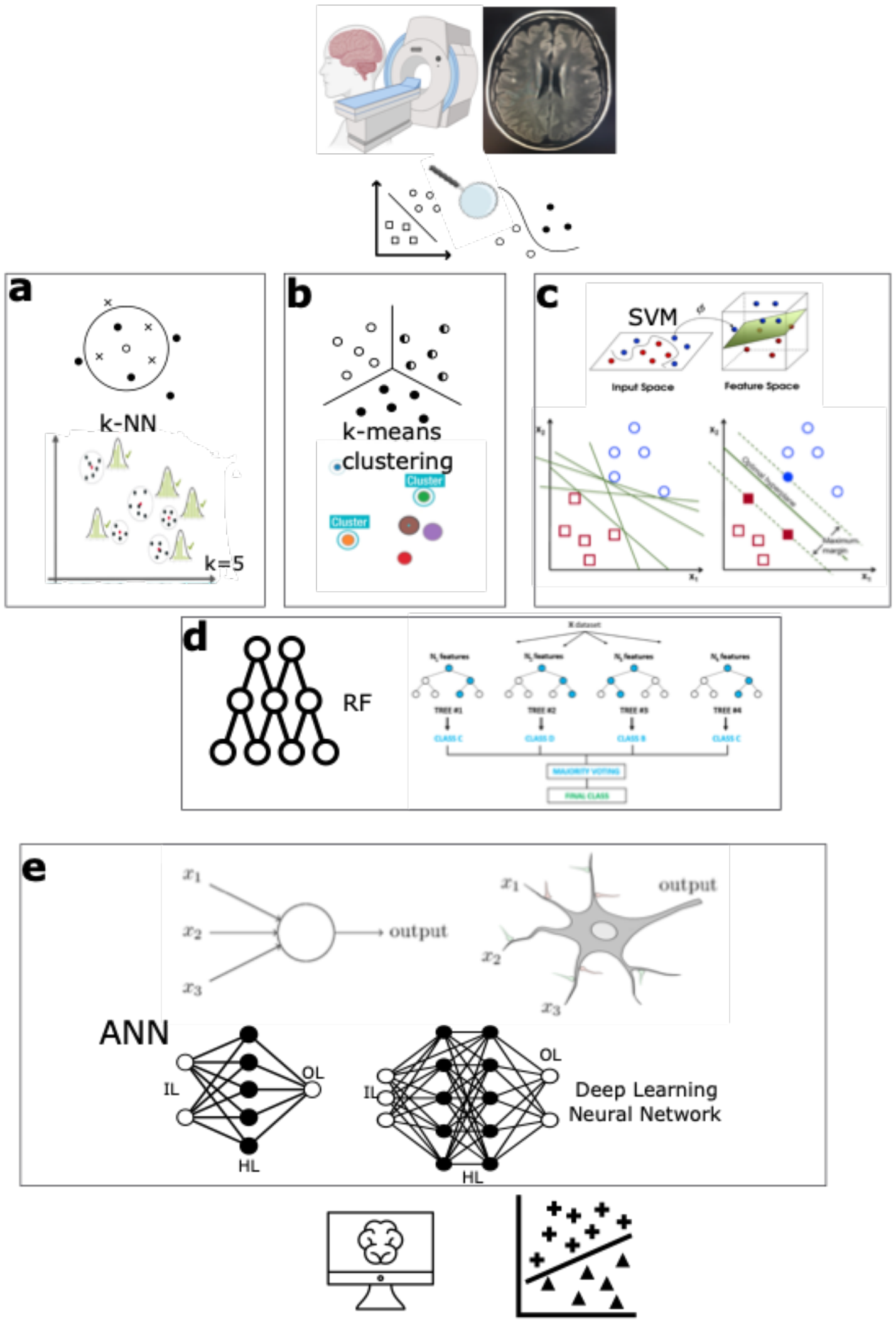

**Fig 8**. Graphical representation of some ML algorithms and the representations of an ANN and DL neural network: (a) correspond to the *k-Nearest Neighbors (k-NN)* algorithm and a representation of k = 5 (the number of nearest neighbors), (b) represents the *k-Means Clustering* algorithm, also is represented a k = 2 clusters, the blue circle in that represents the cluster centroid, (c) is the representation of the *Support Vector Machine (SVM)* algorithm with the optimal separating hyperplane between classes, (d) correspond to *Random Forest (RF)* algorithm and it is represented a forest of classification trees, finally (e) represents the similar concepts used between an artificial neuron and a true neuron with the inputs and outputs, also is presented the architecture of ANN and a DL neural network where IL is the input layers, HL the hidden layers and OL the output layers. This figure was created and adapted using references [197–199].

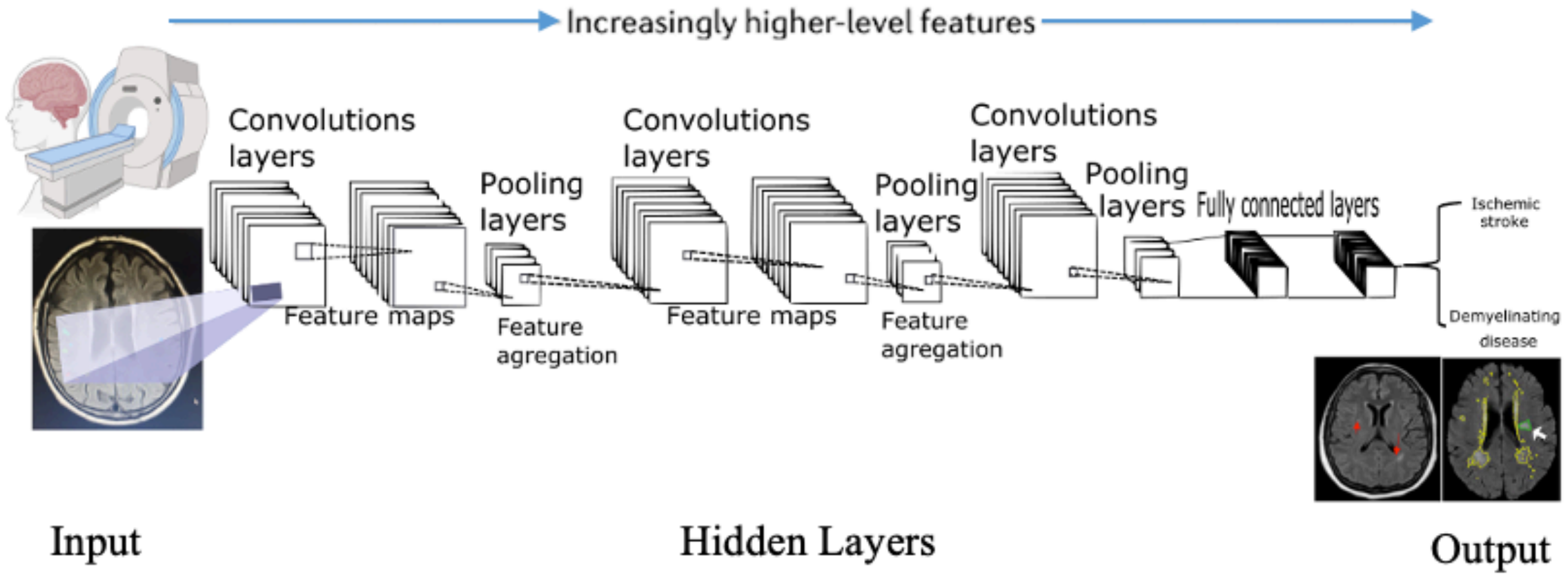

**Fig 9**. Basic architecture of convolutional neural network (CNN), it is showing the convolutional layers that allow get feature maps, the pooling layers for feature aggregation and the fully connected layers for classifications through the global features learned in the previous layers. The level of analysis of features increases with the number of hidden layers. This figure was created and adapted using references [5,120,142,161,164,200].

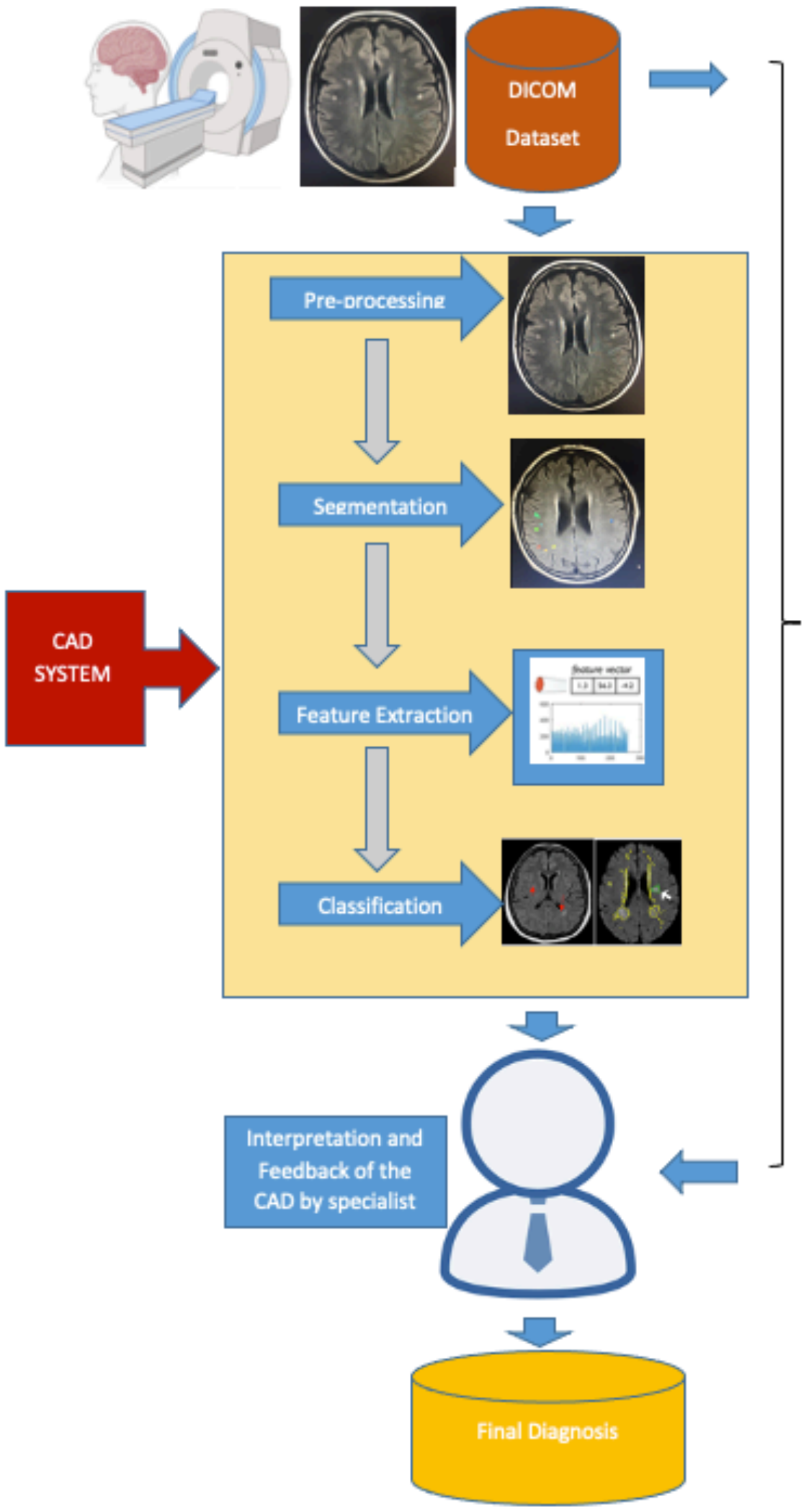

**Fig 10**. Basic structure of CAD system used in the identification of ischemic stroke, in general the stage of feature extraction could be change upon depend the techniques used for extract the feature (ML or DL algorithms)[201]. This figure was created and adapted from[34].

## Tables

**Table 1.** Key Words used in the global semantic structure search

| | |
|---|---|
| **Magnetic Resonance Image** | ( ( ( magnetic* ) AND ( resonanc* ) AND ( imag* OR picture OR visualiz* ) ) OR mri OR mra ) |
| **Processing Brain** | ( algorithm* OR svm OR dwt OR kmeans OR pca OR cnn OR ann ) ) AND ( "deep learning" ) OR ( "neural networks" ) OR ( "machine learning" ) OR ( "convolutional neural network" ) OR ( "radiomics" ) |
| **Disease** | ( ( brain* OR cerebrum ) AND ( ( ischemic AND strok* ) OR ( demyelinating AND ( disease OR "brain lesions" ) ) ) ) |
| **Key words for semantic structure search in Scopus database** | TITLE-ABS-KEY ( ( ( ( magnetic* ) AND ( resonanc* ) AND ( imag* OR picture OR visualiz* ) ) OR mri OR mra ) AND ( ( brain* OR cerebrum ) AND ( ( ischemic AND strok* ) OR ( demyelinating AND ( disease OR "brain lesions" ) ) ) ) AND ( algorithm* OR svm OR dwt OR kmeans OR pca OR cnn OR ann ) ) AND ( "deep learning" ) OR ( "neural networks" ) OR ( "machine learning" ) OR ( "convolutional neural network" ) OR ( "radiomics" ) |

**Table 2.** List of the ten most cited articles according to the normalization of the citations [51]. Also shows the central theme of research, the type of image and the methodology used in the processing.

| Article Title | Author/year | Journal | Total Citations | Norm. Citations | Disease | Type of images/ Dataset | Methodology | Metrics / Observation |
|---|---|---|---|---|---|---|---|---|
| Efficient multi-scale 3D CNN with fully connected CRF for accurate brain lesion segmentation[164] | Kamnitsas, K. et al. (2017) | Medical Image Analysis | 1062 | 7.54 | brain injuries, brain tumors, ischemic stroke | MRI BRATS 2015 ISLES 2015 | 11 layers-deep 3D -CNN | **BRATS2015** DSC: 84.9 Precision: 85.3 Sensitivity: 87.7 **ISLES2015** DSC: 66 Precision: 77 Sensitivity: 63 ASSD: 5.00 Haussdorf: 55.93 |
| ISLES 2015 - A public evaluation benchmark for ischemic stroke lesion segmentation from multispectral MRI[20] | Maier O. et al. (2017) | Medical Image Analysis | 171 | 1.21 | Ischemic stroke | MR- DWI-PWI | RF - CNN | It is a comparison of tools developed in the Challenge ISLES2015 |
| Classifiers for Ischemic Stroke Lesion Segmentation: A Comparison Study[32] | Maier O. et al. (2015) | PLOS ONE | 78 | 2.87 | Ischemic stroke | MRI - Private 37 cases (patients) | Generalized Linear models RDF CNN | DSC[0,1]: 0.80 HD(mm):15.79 ASSD(mm):2.03 Prec.[0,1]:0.73 Rec.[0,1]:0.911 |
| Fully automatic acute ischemic lesion segmentation in DWI using convolutional neural networks[36] | Chen l. et al. (2017) | NeuroImage: Clinical Smart Innovation, Systems and Technologies , vol 105. Springer" | 77 | 0.55 | Ischemic stroke | MR-DWI-Private 741 subjects | CNN DeconvNets: EDD Net MUSCLE Net | DSC: 0.67 Lesion Detection Rate: 0.94 |
| Segmentation of Ischemic Stroke Lesion in Brain MRI Based on Social Group Optimization and Fuzzy-Tsallis Entropy[148] | Rajinikanth, V., Satapathy, S.C.(2018) | Arabian Journal for Science and Engineering | 56 | 2.60 | Ischemic stroke | ISLES 2015 MRI- FLAIR - DWI | Social group optimization monitored Fuzzy-Tsallis entropy | Precision: 98.11% DC:88.54% Sensitivity:99.65% Accuracy:91.17% Specificity: 78.05% |
| Automatic Segmentation of Acute Ischemic Stroke From DWI Using 3-D Fully Convolutional DenseNets[67] | Zhang R. et al. (2018) | IEEE Transactions on Medical Imaging | 47 | 2.18 | Ischemic stroke | MR-DWI - Private 242 subjects: training: 90, testing: 90 Validation: 62 Additional dataset: ISLES2015-SSIS | 3D-CNN | DSC: 79.13% Lesionwise precision: 92.67% Lesionwise F1score: 89.25% |
| ISLES 2016 and 2017-Benchmarking Ischemic Stroke Lesion Outcome Prediction Based on Multispectral MRI[68] | Winzeck S. et al. (2018) | Frontiers in Neurology | 45 | 2.09 | Ischemic stroke | MR- DWI-PWI ISLES 2016 - 2017 | RF CNN | It is a comparison of tools developed in the Challenge ISLES2016-2017 |
| Prediction of tissue outcome and assessment of treatment effect in acute ischemic stroke using deep learning[150] | Nielsen, A. et al. (2018) | Stroke | 42 | 1.95 | Ischemic stroke | MRI - PRIVATE 222 cases (patients) | CNN deep | AUC=0.88±0.12 |
| Enhancing interpretability of automatically extracted machine learning features: application to a RBM-Random Forest system on brain lesion segmentation[165] | Pereira, S. et al.(2018) | Medical Image Analysis | 30 | 1.39 | Brain lesions: Brain tumor ischemic stroke | MRI - Multimodality BRATS 2013: training (30), Leaderboard (25), challenge (10) SPES from MICCAI - ISLES: 30 training, 20 challenge. | Restricted Boltzmann Machine for unsupervised feature learning, and a Random Forest classifier | BRATS2013 DICE: 0.81 SPES DICE: 0.75 ± 0.14 ASSD: 2.43 ± 1.93 |
| Predicting final extent of ischemic infarction using artificial Neural network analysis of Multi-Parametric mri in patients with stroke[149] | Bagher-Ebadian, H. et al. (2011) | Plos ONE | 30 | 1.18 | Ischemic stroke | MRI - Diffusion-weighted-DWI 12 subjects | ANN | Map of Prediction correlation: (r = 0.80, p<0.0001) |

**MRI:** magnetic resonance imaging, **DWI:** diffusion weighted imaging, **PWI:** perfusion-weighted imaging, **FLAIR:** fluid attenuated inversion recovery, **BRATS:** brain tumor image segmentation, **ISLES:** ischemic stroke lesion segmentation, **MICCAI:** medical image computing and computer assisted intervention, **SPES:** stroke perfusion estimation, **ANN:** artificial neural networks, **CNN:** convolutional neural networks, **RF:** random forest**, RDF:** random decision forests, **EDD Net:** ensemble of two DeconvNets, **MUSCLE Net:** multi-scale convolutional label evaluation, **DSC:** dice Score coefficient, **ASSD:** average symmetric surface distance, **HD:** Haussdorf distance.

**Table 3.** List of datasets, details, type (public or private) , web site and their reference dedicated to ischemia (stroke) and demyelinating diseases (MS), also are listed platforms where is possible find datasets for brain medical image processing.

| **Dataset Name** | **Details** | **Type: Private Public** | **web site** | **Reference/Support** |
|---|---|---|---|---|
| MICCAI | MS Lesion Segmentation Challenge: data for competition in order to compare algorithms to segment the MS lesions since 2008 | Public and Private. Some data require subscription | https:// www.nitrc.org/projects/msseg | Akkus et al. (2017)[7] |
| BRATS | Brain Tumor Segmentation: MRI dataset for challenge BRATS since 2012.<br>MRI modalities: T1, T1C, T2, FLAIR. BRATS 2015 contains 220 brains with high-grade and 54 brains with low grade gliomas for training and 53 brains with mixed high- and low-grade gliomas for testing. | Public | https://ipp.cbica.upenn.edu/ | Menze et al. (2015) [202] |
| ISLES | Dataset used for challenge Ischemic Stroke Lesion Segmentation since 2015 in order to **evaluate stroke lesion/ clinical outcome prediction** from acute MRI scans.<br>ISLES has two categories with individual datasets:<br>**SISS:** sub-acute ischemic stroke lesion segmentation, contains 36 subjects with modalities FLAIR, DWI, T2 TSE (Turbo Spin Echo), and T1 TFE (Turbo Field Echo).<br>**SPES:** acute stroke outcome/penumbra estimation, contain 20 subjects with 7 modalities namely: CBF (Cerebral blood flow), CBV (cerebral blood volume), DWI, T1c, T2, Tmax and TTP (time to peak). | Public, require registration and approbation | https://www.smir.ch/ISLES/Start2016 | Maier et al. (2017)[20]<br>Winzeck et al. (2018) [68] |
| MTOP | Mild traumatic Brain Injury Outcome: MRI data for challenge: 27 subjects | Public | https://www.smir.ch/MTOP/Start2016 | Akkus et al. (2017) [7] |
| MSSEG | MS data for evaluating state-of- the-art and advanced segmentation methods. 53 datasets (15 training data and 38 testing data).<br>Modalities: 3D FLAIR, 3D T1-w 3D, T1-w GADO, 2D DP/T2 | Public | https://portal.fli-iam.irisa.fr/msseg-challenge/data. | Commowick et al. (2018)[129] |
| NeoBrainS12 | Set includes T1 and T2 MRI of five infants. Challenge is to compare algorithms for segmentation of neonatal brain tissues and measurement of corresponding volumes using T1 and T2 MRI scans of the brain | Private | https://neobrains12.isi.uu.nl/?page_id=52 | Isgum et al. (2015)[203] |
| MRBrainS | Challenge for segmenting brain structures in MRI scans | Private - Public, require registration and approbation | https://mrbrains13.isi.uu.nl/downloads/ | Mendrik, A.M. et al. (2015)[204] |
| OpenNeuro | A open repository of MRI, MEG, EEG, iEEG, and ECoG datasets | Public | https://openneuro.org/ | ljaf (laura and john Arnold foundation)<br>NSF (National Science Foundation)<br>NIH(National institute on Drug Abuse)<br>Stanford<br>SquishyMedia |
| UK Biobank | International platform of health data resources. Contain MR images from 15000 participants, aiming to reach 100000. | Private - Public, require registration and approbation | https://www.ukbiobank.ac.uk/ | UK Biobank |
| ATLAS | Anatomical Tracings of Lesions After Stroke (ATLAS), is an open-source dataset of 304 T1-weighted MRIs with manually segmented lesions and metadata. | Public, require registration and approbation | https://www.icpsr.umich.edu/web/pages/ | Liew et al.[128] |
| ADNI | Alzheimer's Disease Neuroimaging Initiative, contain data from different type (Clinical, Genetic, MRI image, PET Image, Bioespecimen) | Public, require registration and approbation | http://adni.loni.usc.edu/data-samples/data-types/ | Alzheimers Disease Neuroimaging Initiative (ADNI) |
| ABIDE | Neuroimaging data from the Autism Brain Imaging Data Exchange (ABIDE): 112 datasets from 539 individuals suffering from ASD and 573 typical controls. | Public, require registration and approbation | http://preprocessed-connectomes-project.org/abide/ | Craddock, C. et al. (2013) [205] |
| NIF | Neuroscience information framework project: is a semantically-enhanced search engine of neuroscience information. Data and biomedical resources. | Public | https://neuinfo.org/ | Neuroscience Information Framework (NIF) |
| NEUROVAULT | A public repository of unthresholded statistical maps, parcellations, and atlases of the brain | Public | https://neurovault.org | Gorgolewski et al. (2016) [206] |
| Integrated Datasets | Is a virtual database currently indexing a variety of data sets | Public, require registration | https://scicrunch.org/scicrunch/Resources/record/nlx_144509-1/SCR_010503/resolver | FAIR Data Informatics Lab University of California, San Diego- USA |
| TCIA | Cancer Imaging Archive is a repository with different collections of imaging dataset and diseases | Public, require registration | https://www.cancerimagingarchive.net/collections/ | Department of Biomedical Informatics at the University of Arkansas for Medical Sciences- USA |
| NiftyNEt | An open-source convolutional neural networks platform for medical image analysis and image-guided therapy | Public | https://niftynet.io | Gibson et al. (2018) [207] |
| MONAI | MONAI is a PyTorch-based framework for deep learning in healthcare imaging. It provides domain-optimized foundational capabilities for developing healthcare imaging training workflows in a native PyTorch paradigm. | Public | https://monai.io/ | Project started by NVIDIA & King's College London for AI research community. |
| MSD | Medical Segmentation Decathlon: Challenge of machine learning algorithms for segmentation task. Give a data for 10 tasks: Brain tumor, cardiac, liver, hippocampus, prostate, lung, pancreas, Hepatic Vessel, Spleen, Colon. Modality: MRI and CT | Public | http://medicaldecathlon.com | Simpson et al. (2019)[208] |
| Grand Challenge | A platform for end-to-end development of machine learning solutions in biomedical imaging. | Public-requiere Registration | https://grand-challenge.org/ | Contributors:<br>Bram van Ginneken,<br>Sjoerd Kerkstra, and James Meakin.<br>Radboud University Medical Center in Nijmegen, The Netherlands. |
| StudierFenster | Open Science Platform for Medical Image Processing | Public | http://studierfenster.icg.tugraz.at | TU and the MedUni Graz in Austria |

**Table 4.** List of selected documents related with ischemic stroke and demyelinating disease. It shows their central theme of research, the type of image, the technique used in the processing, the name of dataset and the metrics.

| Author/ Year | Dataset | | | Acquistion Equipment | Research Task | Technique AI | Metrics |
|---|---|---|---|---|---|---|---|
| | Type Access | Image Modality | Composition | | | | |
| Huang et al. (2011) [65] | Private: 36 subjects experiment (rat), three groups of 12 | MRI: T2 + MR -CBF -ADC | Method 1 Training: 1 Testing: 11 Method 2 Training: 11 Testing: 1 | ….. | Stroke | SVM + ANN | ADC+CBF: 86 +- 2.7% 89+-1.4% 93+-0.8% |
| Giese et al. (2020) | MRI-Genetics Interface Exploration (MRI - GENIE) | MRI: FLAIR | 2,529 patients scans | | Stroke | DL | |
| Wu et al. (2019) [59] | MRI-Genetics Interface Exploration (MRI - GENIE) | | 2770 patients scans | | Stroke | 3D CNN | DICE SCORE: 0.81-0.96 |
| Nazari-Farsani et al. (2020) [27] | Private: 192, 3D MRI images | MRI: DWI and ADC | 106 Stroke 86: healthy cases | | Stroke | SVM: linear kernel and cross validation | Accuracy: 73% Precision: 77% Sensitivity: 84% Specificity: 69% |
| Anbumozhi (2020) [15] | Isles 2017: 75 images | MRI: | 52: Healthy 23: stroke | | Stroke | SVM and K-means clustering | Accuracy: 99.8% Precision: 97.3% Sensitivity: 98.8% Specificity: 94% |
| Subudhi et al. (2020) [22] | Private: 192 MRI images | MRI: DWI | 122 PACS 36 LACS, 34 TACS | EM algorithm expectation-maximization (EM) algorithm FODPSO: The FODPSO is an advanced optimization method of Darwinian PSO, | Stroke: LACS, PACS, TACS | SVM and Random Forest (RF) | Accuracy: 93.4% DSC: 0.94 |
| Qiu et al. (2020) [151] | PRIVATE_ 1000 patients | MRI and CT | | Manually defined features U-net Transfer Learning | Stroke | Random Forest | Accuracy: 95% |
| Li et al. (2004) [152] | Private: 20 patients | DT MRI: (difusion tensor MR) | 20 | Manual Lesion tracing | Stroke | MSSC+PVVR | Similarity: 0.97 |
| Wang et al. (2013) [153] | 20 patients | | | Level set framework, based in local region models | | LLC (Local Linearl Classifier | Segmentation Error Rate: 0.0825 |
| Raina et al.(2020) [154] | ISLES 2015 | FLAIR, DWI, T1, T1-contrast | 28 patients | Quasi-symmetry | Stroke | TwoPathCNN+NLSymm Wider2dSeg+NLSymm | Dice: 0.54; 0.62 Precision: 0.52; 0.68 Recall: 0.65; 0.60 |
| Grosser et al. (2020) [155] | Private: 99 Patients | MRI: DWI and PWI, FLAIR | 99 patients | Software AnToNIa: tool for analysis multipurpose MR perfusion | Stroke | ML: Logistic Regression, random forest and XGBoost | ROC AUC: 0.893*/-0.085 |
| | | | | datasets | | | |
| Lee et al. (2020) [158] | Private: 355 patients | MRI: DWI-FLAIR | 355 patients: 299 train; 56 to test | 89 vector features | Stroke | ML: Logistic Regression, Landom forest and SVM | Sensitivity and Specificity; Logistic Regression: 75.8% , 82.6% SVM: 72.7%, 82.6%. Random Forest: 75.8%,82.6% |
| Melingi y Vivekanan d(2018) [131] | Private, real time dataset | MR DWI | .. | .. | Ischemic stroke | Kernelized Fuzzy C-means (KFCM) clustering and SVM | Accuracy:98.8% Sensitivity: 99% |
| Clêrigues et al. (2020) [156] | ISLES 2015 | multimodal MR imaging T1, T2, FLAIR, DWI | sub-acute ischemic stroke segmentation (SISS): 28 training and 36 testing cases stroke penumbra estimation sub-task (SPES): 30 training and 20 testing cases | | SISS and SPES | U-Net based CNN architecture using 3D convolutions, 4 resolution steps and 32 base filters | SISS: DSC=0.59 ± 0.31 SPES: DSC=0.84 ± 0.10 |
| Kumar et al. (2020) [136] | ISLES 2015<br>ISLES 2017 | Multimodal MR imaging<br>T1,T2,FLAIR, DWI | ISLES 2015<br>(SISS): 28 training and 36 testing cases<br>(SPES): 30 training and 20 testing cases<br>ISLES 2017:  43 training and 32 test cases | | Ischemic Stroke | Classifier-Segmenter Network: A combination of U-Net and Fractal Networks | ISLES 2015 - SISS<br>Accuracy: 0.9914<br>Dice-Coeff:0.8833<br>Recall: 0.8973<br>Precision: 0.8760<br><br>ISLES 2015 - SPES<br>Accuracy: 0.9908<br>Dice-Coeff: 0.8993<br>Recall: 0.9091<br>Precision: 0.9084<br><br>ISLES 2017<br>Accuracy: 0.9923<br>Dice-Coeff: 0.6068<br>Recall: 0.6611<br>Precision: 0.6141 |
| Leite et al. (2015) [13] | Private images from 77 patients: | T2-weighted MRI | 50 with MS,<br>19 heatlhy,<br>4 stroke<br>75% trainining set<br>25% testing set | | WMH<br>Stroke<br>MS | Classifiers:<br>SVM with RBF kernel<br>OPF: optimum path forest with decision tree,<br>LDA: Linear discriminant analysis with PCA<br>kNN:k-nearest neighbor, K=1, K=3,k=5 | Accuracy for classifier demyelinating and ischemic<br>SVM: 86.5 +/- 0.09<br>kNN: 83.57  0.07<br>OPF:81.23  0.06<br>LDA: 83.52  0.06 |
| Ghafoorian et al. (2016) [14] | Private: 362 MRI Scans patients | FLAIR | 312 Training<br>50 Testing | | WM<br>Cerebral<br>SVD | AdaBoost + Random Forest | Free-response receiving operating characteristic (FROC)  analysis<br>Sensitivity: 0.73 with 28 false positives |
| Bowles et al. (2017) [137] | Private Brain Research Imaging Centre of Edinburgh 127 Subjects | FLAIR | 20 Training | GE Signa Horizon HDx 1.5 T clinical scanner | Cerebral SVD MS | Image synthesis Gaussian Mixture models SVM all of them compared with publicly available methods LST LesionTOADS and LPA | DSC: 0.70 ASSD: 1.23 HD:38.6 Prec:0.763 Recall:0.695 Fazekas Correlation: 0.862 ICC:0.985 |
| Brosch et al. (2016) [159] | MICCAI 2008<br>ISBI 2015<br>PRIVATE-CLINICAL: 377 subjects | T1, T2, PD and FLAIR<br>T1, T2, PD and FLAIR<br>T1, T2 and FLAIR | MICCAI<br>Training: 20<br>Testing: 23<br>ISBI<br>Training: 20<br>Testing: 61<br>Validation: 1<br>CLINICAL:<br>Training: 250<br>Testing: 77<br>Validation: 50 | Private: 1.5T and 3T scanners | MS | 3D Convolutional Encoder Networks | MICCAI<br>VD: UNC 63.5%, CHB 52.0%<br>TPR: UNC 47.1%, CHB 56.0%<br>FPR: UNC 52.7%, CHB 49.8%<br>ISBI<br>DSC: 68.3%<br>LTPR: 78.3%<br>LFPR: 64.5%<br>CLINICAL:<br>DSC: 63.8%<br>LTPR: 62.5%<br>LFPR: 36.2%<br>VD: 32.9 |
| Valverde et al. (2017) [141] | (a) MICCAI 2008<br>(b) Private, Clinical 1: 35 subjects<br>(c) Private, Clinical 2: 25 subjects | (a) T1, T2, PD and FLAIR<br>(b)T1, T2 and FLAIR<br>(c)T1, T2 and FLAIR | (a)<br>Training:20<br>Testing: 25<br>(b)<br>Training: ...<br>Testing: ..<br>(a)<br>Training:..<br>Testing: ... | … | MS | 3D Cascade CNN | (a)<br>VD: UNC 62.5%, CHB 48.8%<br>TPR: UNC 55.5%, CHB 68.7%<br>FPR: UNC 46.8%, CHB 46.0%<br>(b)<br>DSC: 53.5%<br>VD:30.8%<br>TPR: 77.0%<br>FPR: 30.5%<br>PPV:70.3%<br>(c):<br>DSC: 56.0%<br>VD:27.5%<br>TPR: 68.2%<br>FPR: 33.6%<br>PPV:66.1% |
| Praveen et al. (2018) [143] | ISLES 2015: 28 volumetric brains | T1, FLAIR, DWI, and T2. | Training: 27<br>Testing: 1 | .. | Ischemic stroke | Stacked sparse autoencoder (SSAE)<br>+SVM | Precision: 0.968<br>DC:0.943<br>Recall:0.924<br>Accuracy:0.904 |
| Guerrero et al. (2018) [161] | Private Brain Research Imaging Centre of Edinburgh 127 Subjects | T1 and FLAIR | Training:127 | GE Signa Horizon HDx 1.5 T clinical scanner (General Electric, Milwaukee, WI), | WMH<br>Stroke | CNN - u-shaped residual network (uResNet) architecture | WMH Dice(std): 69.5(16.1)<br>Stroke Dice (std): 40.0(25.2) |
| Mitra et al. (2014) [139] | Private 36 patients | T1W, T2W, FLAIR and DWI | | 3T MR (Magneton Trio; Siemens, Erlangen, Germany) scanner | Lesion (WMH) Ischemic stroke MS | Random Forest | DSC(std): 0.60(0.12) PPV(std): 0.75(0.18) TPR(std): 0.53(0.13) SMAD(std):3.06(3.17) VD(%)(std):32.32(21.64) |
| Menze et al. (2016) [30] | BRATS 2012-2013 Glioma Data Private dataset for Stroke(Zurich): 18 datasets | T1, T2, T1c and FLAIR | … | .. | Glioma Ischemic Stroke | Gaussian mixtures and a probabilistic tissue atlas that employs expectation-maximization (EM) segmenter | FLAIR BRATS glioma, DSC: 0.79 (±0.06) Zurich Stroke DSC: 0.79 (±0.07) T1c BRATS glioma, DSC: 0.66 (±0.14) Zurich Stroke DSC: 0.6479 (±0.18) |
| Subudhi et al. (2018) [209] | Private:142 patients | MR DWI | | GE Medical Systems MRI 1.5 T and flip angle of 55° | Ischemic Stroke | Watershed, relative fuzzy connectedness, and guided filter + MLP (multilayer perceptron) and Random Forest | MLP NN DSC: 0.86 Random Forest DSC: 0.95 |
| Boldsen et al. (2018) [157] | PRIVATE: 108 patients | MRI: DWI | .. | GE Signa Excite 1.5T, GE Signa Excite 3T, GE Signa HDx 1.5T, GE Signa Horizon 1.5T, Milwaukee, WI; Siemens TrioTim 3T, Siemens Avanto 1.5T, Siemens Sonata 1.5T, Erlangen, Germany; Philips Gyroscan NT 1.5T, Phillips Achieva 1.5T, and Philips Intera 1.5T, Best, Netherlands | Ischemic Stroke | ATLAS machine learning Algorithm COMBAT Stroke | ATLAS DICE: 0.6122 COMBAT Stroke DICE:0.5636 |
| Karthik et al. (2019) [132] | ISLES 2015: 28 volumetric brains | Multimodal MRI T1,T2,FLAIR, DWI | .. | .. | Ischemic stroke | FCN, Propose variant of U-Net architecture CNN | DSC: 0.70 |
| Wottschel et al. (2019) [162] | Private: 400 patients multicentre | T1, T2-weighted MRI | .. | .. | MS-WM lesions | SVM with Cross-Validation | Accuracy for multicentre (all dataset): 64.8%-70.8% Accuracy for individual center (small dataset) 64.9%-92.9% |

**CBF:** Cerebral blood flow, **ADC:** apparent diffusion coefficient, **SVD:** small vessel disease, **TACS:** total anterior circulation stroke syndrome, **PACS:** partial anterior circulation stroke syndrome, **LACS:** lacunar stroke syndrome, **SISS:** sub-acute ischemic stroke segmentation, **SPES:** stroke penumbra estimation

**Table 5**. Summary of the CNN architectures and principal libraries used for to build models of DL. The data for this was collected from [76,210,211] .

| ARCHITECTURES OF CNN | | |
|---|---|---|
| Name | **Reference** | **Details** |
| LeNet | Backpropagation Applied to Handwritten Zip Code Recognition[212] | Yann LeCun 1990. Read zip codes, digits. |
| AlexNet | ImageNet Classification with Deep Convolutional[213] Neural Networks | Alex Krizhevsky, Ilya Sutskever and Geoff Hinton, in 2012 ILSVRC challenge. Similar to LeNet but deeper, bigger and featured Convolutional |
| ZF Net | Visualizing and Understanding Convolutional Networks[214] | Matthew Zeiler and Rob Fergus, ILSVRC,2013. Expanding the size of the middle convolutional layers and making the stride and filter size on the first layer smaller |
| GoogLeNet | Going Deeper with Convolutions[215] | Szegedy et al., from Google, ILSVRC 2014. Going Deeper with Convolutions |
| VGGNet | Very Deep Convolutional Networks for Large-Scale Image Recognition[216] | Karen Simonyan and Andrew Zisserman in ILSVRC 2014. The depth of the network is a critical component for good performance |
| ResNet | Residual Network[166] | Kaiming He et al., in ILSVRC 2015. It features special skip connections and a heavy use of batch normalization. |
| Highway nets | Highway Networks[217] | The architecture is characterized by the use of gating units which learn to regulate the flow of information through a network. |
| DenseNet | Densely Connected Convolutional Networks[218] | Dense Convolutional Network (DenseNet) connects each layer to every other layer in a feed-forward fashion. |
| SENets | Squeeze-and-Excitation Networks[219] | Model interdependencies between channels of the relationship of features used in the traditional CNN. |
| NASNet | Neural Architecture Search Network[220] | Authors propose to search an architectural building block on a small dataset and then transfer the block to a larger dataset. |
| YOLO | You Only Look Once[221] | A unified model for object detection |
| GANs | Generative Adversarial Nets[104] | Framework for estimating generative models via adversarial nets |
| Siamese nets | Siamese Networks[222] | A Siamese Neural Network is a class of neural network architectures that contain two or more identical subnetworks. 'identical' here means, they have the same configuration with the same parameters and weights. |
| U-Net | U-Net: Convolutional Networks for Biomedical Image Segmentation[145] | The architecture consists of a contracting path to capture context and a symmetric expanding path that enables precise localization. |
| V-net | V-Net: Fully Convolutional Neural Networks for Volumetric Medical Image Segmentation[223] | Architecture for 3D image segmentation based on a volumetric, fully convolutional, neural network |
| | **LIBRARIES USED TO BUILD DL MODELS** | |
| GIMIAS | http://www.gimias.org/ | A workflow-oriented environment for solving advanced, biomedical image computing and individualized simulation problems |
| SPM | https://www.fil.ion.ucl.ac.uk/spm/ | The analysis of brain imaging data sequences using Statistical Parametric Mapping as an assessment of spatially extended statistical processes used to test hypotheses about functional imaging data |
| FSL | https://fsl.fmrib.ox.ac.uk | A collection of analysis tools for FMRI, MRI and DTI brain imaging data |
| PyBrain | http://pybrain.org/ | Reinforcement Learning, Artificial Intelligence and Neural Network Library |
| Caffe | http://caffe.berkeleyvision.org/ | A deep ML framework |
| PyMVPA | http://www.pymvpa.org/ | Statistical learning analysis platform |
| Weka | https://www.cs.waikato.ac.nz/ml/weka/ | Data mining platform |
| Shogun | http://www.shogun-toolbox.org/ | Machine learning framework |
| SciKit Learn | http://scikit-learn.org | Scientific computation libraries |
| PRoNTo | http://www.mlnl.cs.ucl.ac.uk/pronto/ | Machine learning framework |
| Tensorflow | http://playground.tensorflow.org | Created by Google. It provides excellent performance and multiple CPUs and GPUs support |
| Theano | https://pypi.org/project/Theano/ | Easy to build a network but challenging to create a full solution. Uses symbolic logic and written in Python |
| Keras | https://keras.io | Created in Python, it is possible to use with Theano or Tensorflow backend |
| Torch | http://torch.ch/docs/tutorials-demos.html | Created in C. Performance is very good |
| Pytorch | https://pytorch.org | It is a Python front end to the Torch computational engine. It is an integration of Python with the Torch engine. Performance is higher than Torch with GPU integration facility. |